\documentclass[preprint2]{aastex}

\usepackage{color}
\usepackage{hyperref}
\usepackage{amsmath}
\slugcomment{}
\shorttitle{Hydrogenated PAHs}
\shortauthors{Steglich et al.}

\begin{document}


\title{The abundances of hydrocarbon functional groups in the interstellar medium inferred from laboratory spectra of hydrogenated and methylated polycyclic aromatic hydrocarbons \linebreak \textit{\small(published in Astrophysical Journal Supplement Series, 208 (2013) 26)}}


\author{M. Steglich\altaffilmark{1}, C. J\"ager, F. Huisken}
\affil{Laboratory Astrophysics Group of the Max Planck Institute for Astronomy at the Friedrich Schiller University Jena, Institute of Solid State Physics, Helmholtzweg 3, D-07743 Jena, Germany}
\email{M.Steglich@web.de}

\author{M. Friedrich, W. Plass}
\affil{Institute of Inorganic and Analytical Chemistry, Friedrich Schiller University Jena, Humboldtstra\ss e 8, D-07743 Jena, Germany}

\author{H.-J. R\"ader, K. M\"ullen}
\affil{Max Planck Institute for Polymer Research, Ackermannweg 10, D-55128 Mainz, Germany}

\and

\author{Th. Henning}
\affil{Max Planck Institute for Astronomy, K\"onigstuhl 17, D-69117 Heidelberg, Germany}


\altaffiltext{1}{present address: Department of Chemistry, University of Basel, Klingelbergstrasse 80, CH-4056 Basel, Switzerland}


\begin{abstract}
Infrared (IR) absorption spectra of individual polycyclic aromatic hydrocarbons (PAHs) containing methyl (\sbond CH$_3$), methylene (\includegraphics[scale=0.18]{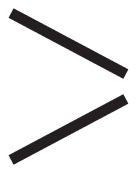}CH$_2$), or diamond-like \includegraphics[scale=0.18]{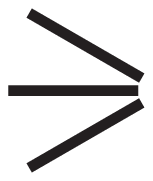}CH groups and IR spectra of mixtures of methylated and hydrogenated PAHs prepared by gas phase condensation were measured at room temperature (as grains in pellets) and at low temperature (isolated in Ne matrices). In addition, the PAH blends were subjected to an in-depth molecular structure analysis by means of high-performance liquid chromatography (HPLC), nuclear magnetic resonance (NMR) spectroscopy, and matrix-assisted laser desorption/ionization time-of-flight mass spectrometry (MALDI-TOF). Supported by calculations at the density functional theory (DFT) level, the laboratory results were applied to analyze in detail the aliphatic absorption complex of the diffuse interstellar medium (ISM) at 3.4 $\mu$m and to determine the abundances of hydrocarbon functional groups. Assuming that the PAHs are mainly locked in grains, aliphatic CH$_x$ groups ($x$ = 1,2,3) would contribute approximately in equal quantities to the 3.4 $\mu$m feature ($N_{\text{CH}x}$ $/$ $N_\text{H}$ $\approx$ 10$^{-5}$ $-$ 2 $\times$ 10$^{-5}$). The abundances, however, may be two to four times lower if a major contribution to the 3.4 $\mu$m feature comes from molecules in the gas phase. Aromatic \includegraphics[scale=0.18]{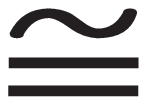}CH groups seem to be almost absent from some lines of sight, but can be nearly as abundant as each of the aliphatic components in other directions ($N_{\text{\includegraphics[scale=0.12]{arom-bond.eps}CH}}$ $/$ $N_\text{H}$ $\lesssim$ 2 $\times$ 10$^{-5}$; upper value for grains). Due to comparatively low binding energies, astronomical IR emission sources do not display such heavy excess hydrogenation. At best, especially in proto-planetary nebulae, \includegraphics[scale=0.18]{doublebond.eps}CH$_2$ groups bound to aromatic molecules, i.e., excess hydrogens on the molecular periphery only, can survive the presence of a nearby star.
\end{abstract}

\keywords{dust, extinction --- ISM: abundances --- ISM: lines and bands --- ISM: molecules --- methods: laboratory --- molecular data}

\section{Introduction} \label{sec_intro}
The formation of carbon-containing molecules and grains in the envelopes and outflows of carbon-rich stars is studied in the laboratory using techniques involving laser ablation of graphite and laser-induced pyrolysis of a hydrocarbon precursor gas. The products of these laser-based syntheses are considered as analogs of cosmic molecules and grains \citep[e.g.,][]{menella99, schnaiter99, duley05, jager06}. Conclusions concerning the nature and structural composition of carbonaceous material in the interstellar medium (ISM) are mainly based on comparisons of astronomical infrared (IR) observations with absorption and emission spectra of laboratory analogs.

Heavily reddened sight lines of the diffuse ISM feature a prominent complex of absorption bands around 3.4 $\mu$m, which has been attributed to the CH stretching vibrations of saturated carbon atoms \citep[e.g.,][]{butchart86, adamson90, sandford91, pendleton94, chiar00}. Based on an analysis of laboratory spectra of several dust analogs with respect to the interstellar IR absorption, \citet{pendleton02} inferred that the organic refractory material in the diffuse ISM is predominantly hydrocarbon in nature, possesses little nitrogen, oxygen, or long alkane chains, and resembles more closely hydrocarbon residues from laser-based syntheses than energetically processed ice residues. From the subpeak intensities of the 3.4 $\mu$m feature it was furthermore concluded that methylene groups (\includegraphics[scale=0.18]{doublebond.eps}CH$_2$) are about fifty per cent more abundant than methyl groups \citep[\sbond CH$_3$;][]{ehrenfreund91}. The additional presence of diamond-like \includegraphics[scale=0.18]{triplebond.eps}CH groups was noted as well \citep{allamandola92}. In order to rationalize the presence of the 3.4 $\mu$m feature, an efficient reformation mechanism for aliphatic CH bonds has to be active to counterbalance the approximated photodestruction rates in the diffuse ISM \citep[e.g.,][]{menella01}.

Bands around 3.4 $\mu$m were also found in IR emission sources, such as proto-planetary and reflection nebulae. However, unlike the diffuse ISM, the features around 3.4 $\mu$m are usually weaker than the one at 3.3 $\mu$m, which is generally attributed to aromatic CH stretching vibrations in polycyclic aromatic hydrocarbons \citep[PAHs;][]{geballe92, joblin96, goto03, tielens08, li12}. The shape of the 3.4 $\mu$m complex seen in emission also differs from that observed in absorption. It was concluded that, compared to methylene groups, the abundances of methyl groups are negligible \citep{bernstein96, wagner00, duley05}.

All previous laboratory studies, which were aiming to reproduce the carriers of the 3.4 $\mu$m absorption complex via laser-based synthesis routes, focused on the production and characterization of nano- to macroscopic dust grains. In addition to absorption spectroscopy, other structure analysis techniques, especially electron microscopy, were sometimes employed \citep[e.g.,][]{bradley05, jager09}. However, the characterization usually did not involve methods able to independently probe the functional groups of a sample, which, in turn, would support the analysis of IR spectra and ease feature assignments. Furthermore, no information regarding the absolute abundances of individual functional groups in the diffuse ISM were obtained from studying grains. 

Alongside nanometer-sized carbon grains, copious amounts of PAHs can be produced in a laser-induced pyrolysis experiment. The electronic absorption spectra of such PAH blends were discussed in an astrophysical context in previous publications \citep{steglich10, steglich12}. It was found that such rich molecular mixtures can contribute to or be entirely responsible for the interstellar ultraviolet (UV) bump at 217.5 nm. A possible connection to the diffuse interstellar bands, as often proposed, could not be verified. Here, we present an in-depth study of these analogs of cosmic molecules with a focus on the CH stretching vibrations of \includegraphics[scale=0.18]{doublebond.eps}CH$_2$ and \sbond CH$_3$ groups. To complement the IR spectroscopic investigations, we carried out a structure analysis of the molecular mixtures by means of high-performance liquid chromatography \citep[HPLC; already in][]{steglich10, steglich12}, nuclear magnetic resonance (NMR) spectroscopy, and matrix-assisted laser desorption/ionization in combination with time-of-flight mass spectrometry (MALDI-TOF). The implications in the context of astrophysics are introduced.

This study comprises the following parts: The MALDI-TOF and NMR experiments along with details about the sample preparation are described in Sect. \ref{sec_structure}. Surprisingly, the NMR studies of the fractionated samples revealed an abundant presence of methyl (\sbond CH$_3$) and methylene (\includegraphics[scale=0.18]{doublebond.eps}CH$_2$) groups attached to the periphery of the aromatic molecules. This finding has direct consequences for the IR absorption spectra. As opposed to the electronic spectra, which were in large part dominated by the aromatic character of the central C atoms, the IR spectra display strong aliphatic characteristics. To support their analysis and to identify characteristic absorption features, we conducted additional experiments on individual hydrogenated and methylated PAH molecules (Sect. \ref{sec_indiv}). These results, in combination with theoretical calculations at the level of density functional theory (DFT), were then used for the determination of absorption cross sections (Sect. \ref{sec_theoryIR}). Section \ref{sec_mixPAHs} is dedicated to the IR spectra of the fractionated PAH blends. Regarding the shapes and relative intensities of characteristic absorption bands, the pyrolysis products offer a better representation of astrophysical PAH distributions than individual molecules. In addition to pellet measurements, low-temperature matrix isolation experiments were performed to obtain the band positions for physical conditions as encountered for isolated molecules in the ISM. The contribution finishes with an application to astrophysics in Sect. \ref{sec_ISM}. Based on the experimental results and an analysis of the interstellar 3.4 $\mu$m absorption complex, we derived the abundances of hydrocarbon functional groups in the diffuse ISM. The conclusions are presented in Sect. \ref{sec_summary}.

\section{Gas-phase condensed PAH mixtures with aromatic and aliphatic structures} \label{sec_structure}
\subsection{Preparation of the PAH mixtures}
The PAH mixtures were prepared by extraction from laser pyrolysis condensates \citep{jager06}. A more detailed description of the condensation technique is provided  in \citet{morjan03}. Briefly, a 60 W continuous wave CO$_2$ laser was applied to decompose the hydrocarbon precursor gas ethene (C$_2$H$_4$) inside a vacuum flow reactor. Additional argon served as buffer gas. The reactor was operated at a pressure of 750 mbar and with flow rates of 40 ml min$^{-1}$ ethene and 1600 ml min$^{-1}$ argon. Molecules and particles formed in the hot (1100 $^\circ$C), flame-like condensation zone and were collected afterward in a filter made of polytetrafluoroethylene. In this study, we were only interested in the soluble molecules, which we extracted from the filter first with methanol and second with dichloromethane (DCM). An additional size separation of the DCM extract was performed via HPLC. Five different samples were studied. The first sample consisted of the combined methanol and DCM extracts and contained all soluble components. The second sample was the DCM extract, which had a similar composition as the first sample except that the low-mass components soluble in methanol had been mostly removed. Finally, three different extracts size-separated by HPLC were analyzed. The three HPLC extracts are named according to the retention times, at which the comprised molecules appeared in the HPLC spectrum \citep[see][]{steglich12}. The time windows are 7 $-$ 20 min, 20 $-$ 35 min, and 7 $-$ 35 min, where the last sample basically contains all molecules from the first two windows. The fraction with retention time 7 $-$ 20 min roughly covers the molecular range from anthanthrene (C$_{22}$H$_{12}$) to ovalene (C$_{32}$H$_{14}$). All larger species should be included in the fraction ``20 $-$ 35 min''. However, as discussed previously \citep{steglich12}, smaller PAHs with very high solubilities, such as anthracene (Ant; C$_{14}$H$_{10}$), phenanthrene (Phn; C$_{14}$H$_{10}$), or pyrene (Pyr; C$_{16}$H$_{10}$), are present as impurities in all samples due to imperfect size separation. In the following, the samples are additionally named according to the number of the filter from which they originate. We prepared four different filters (labeled from ``1'' to ``4''), which each took about 60 hours of pyrolysis runtime to fill up. Due to different storage times and other small variations in the experimental conditions, the sample compositions from the different filters might differ slightly.

Besides HPLC, the various extracts were characterized by gas chromatographic/mass spectrometric (GC/MS) analysis, MALDI-TOF, and NMR spectroscopy. Results of the GC/MS \citep[toluene extract only;][]{jager06, jager07} and HPLC analyses \citep{steglich10, steglich12} were published previously. A detailed sample characterization via NMR and MALDI-TOF spectroscopy is presented below.

\subsection{MALDI-TOF analysis} \label{sec_MALDI}
In order to obtain information about the molecular sizes, four of our samples were subjected to an in-depth MALDI-TOF analysis. The experiments were performed on a Bruker Daltonics Reflex II MALDI-TOF mass spectrometer equipped with a 337 nm nitrogen laser. The instrument was optimized to allow resolution of isotopes up to masses of 6000 amu. Between 200 and 600 amu, the mass resolution $m / \Delta m$ was about 1000 to 1500. Prior to the measurements, the investigated samples were mixed with the matrix trans-2-[3-(4-tert-butylphenyl)-2-methyl-2-propenylidene]malonitrile (DCTB) in solid state in a ball mill (solvent free sample preparation). Through transfer of absorbed laser power, the matrix enables a gentle evaporation of the PAH molecules which are simultaneously ionized by photoionization. After that, the created ions can be detected by a time-of-flight mass spectrometer.

\begin{figure*}\begin{center}
\epsscale{2.15} \plotone{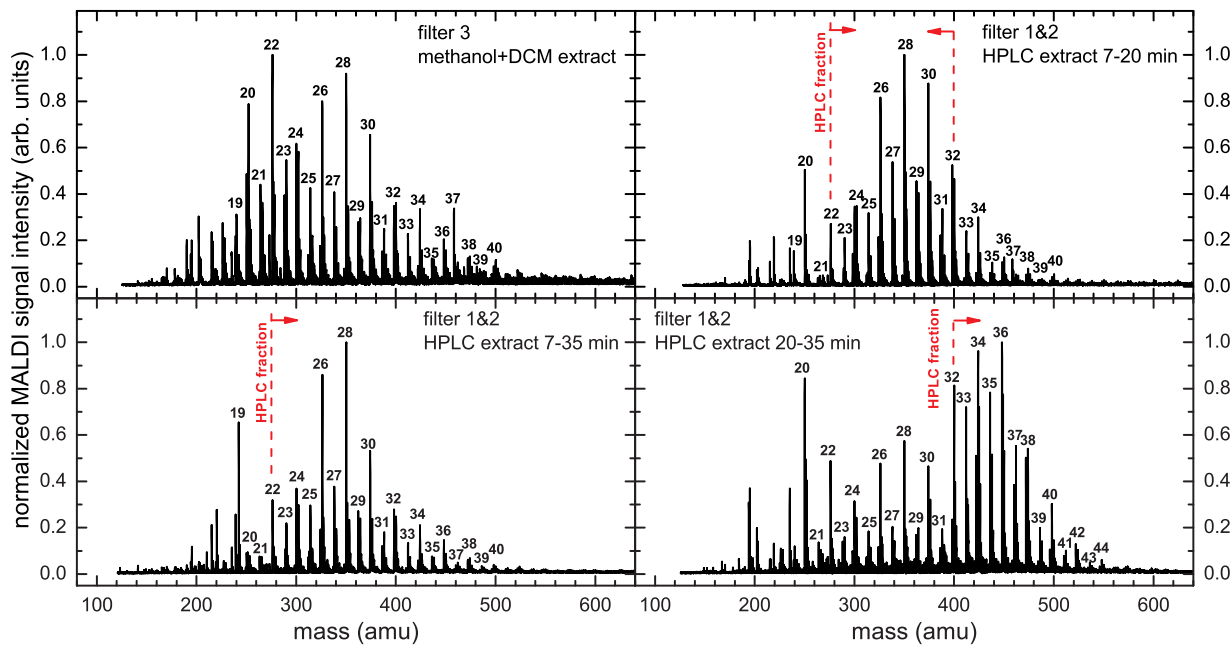} \caption{MALDI-TOF spectra of PAH extracts from the laser pyrolysis. The number of C atoms is indicated at each main peak. For the HPLC extracts, the desired molecular sizes after fractionation are displayed as well (dashed red lines).} \label{fig_MALDIall}
\end{center}\end{figure*}

The MALDI-TOF spectra of the combined methanol/DCM extract and the three HPLC fractions are presented in Fig. \ref{fig_MALDIall}. It is important to mention that purely aliphatic species will not be ionized by MALDI and cannot be traced by this method. To check for the presence of aliphatic molecules, we performed additional experiments utilizing field desorption mass spectrometry which ionizes aliphatic and aromatic molecules as well (not shown here). The thus measured spectra are very similar to the MALDI-TOF spectra, implying that completely saturated hydrocarbons are not abundant components of our samples. According to our experience, the presented mass spectra provide a rather good impression of the overall molecular size distributions of the soluble extracts. The lower desorption probabilities of the larger molecules are somewhat counterbalanced by their higher ionization probabilities, at least for the presented mass range. However, the signal intensities drop considerably as the molecular sizes increase and even larger PAHs are exceedingly difficult to detect. All four samples clearly contain PAHs composed of up to about 44 C atoms. The undissolved soot from the laser pyrolysis also contains molecules with more than 100 or even 200 C atoms \citep[about 3000 amu; see][]{jager09}. Because of extremely low solubilities, however, such large species are at best present as traces in our extracts.

Comparing the four spectra in Fig. \ref{fig_MALDIall}, we can assess the extent of the HPLC fractionation procedure. The combined methanol/DCM extract contains a broad distribution of molecules, ranging in size from 14 to 42 C atoms. Weak peaks at masses smaller than 178 amu are mainly due to the matrix. By comparison, the size distribution of the HPLC extract ``7 $-$ 20 min'' is narrower as molecules with less than 22 and more than 32 C atoms have been partly removed. Obviously, the size separation is not perfect and PAHs outside of the fractionation window are still present. At first glance, the mass spectrum of the sample ``7 $-$ 35 min'' is very similar, but in detail the peak heights at certain masses are different. For instance, the peak at the mass of C$_{19}$H$_{14}$, corresponding to several PAHs with one methyl group, is comparatively strong while the peak intensities at 20 and 30 C atoms have decreased. The mass spectrum of the last extract, ``20 $-$ 35 min'', displays an accumulation of strong peaks at high masses between 398 and 550 amu corresponding to 32 to 44 C atoms. However, molecules as small as 178 amu (C$_{14}$H$_{10}$) are still present.

\begin{figure*}\begin{center}
\epsscale{2.15} \plotone{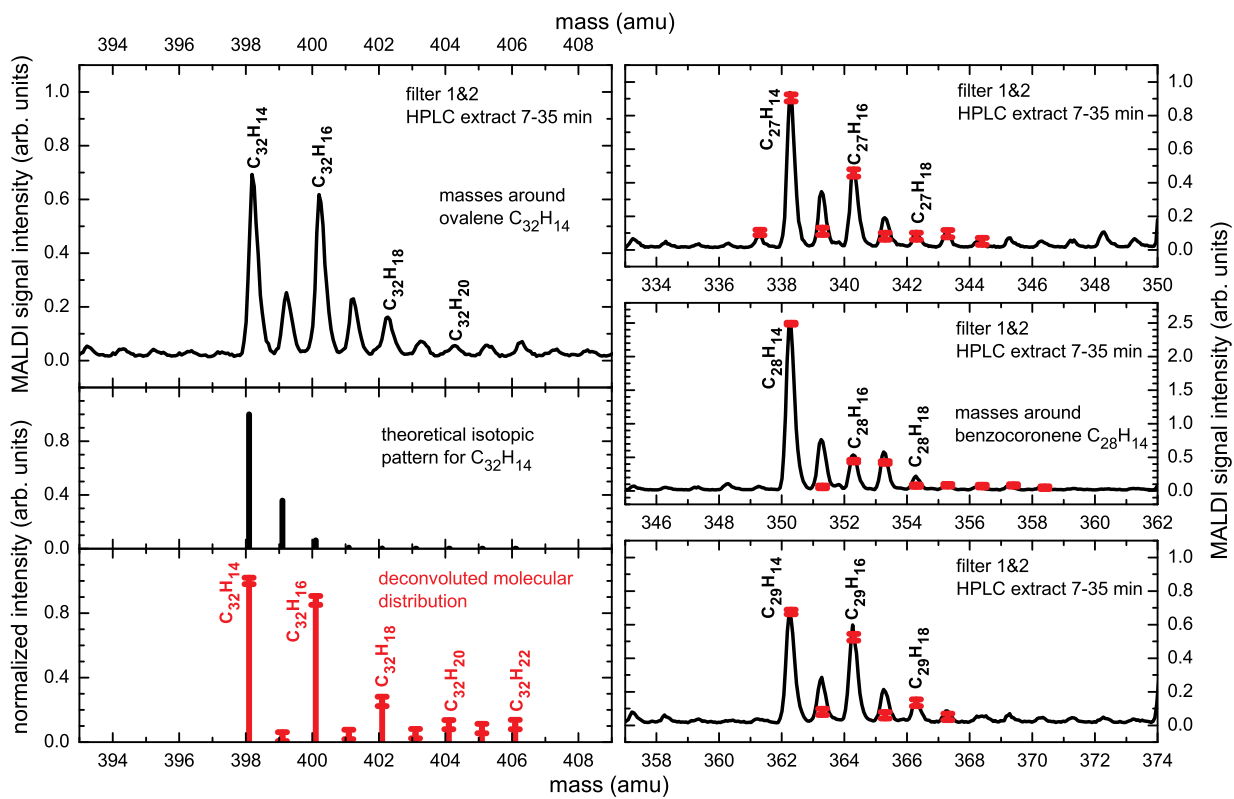} \caption{MALDI-TOF mass spectra of the HPLC extract ``7 $-$ 35 min''. Left panels: Masses around C$_{32}$H$_{14}$. The measured mass spectrum (top) was corrected for isotopes to obtain the deconvoluted molecular distribution (bottom). The isotopic pattern of C$_{32}$H$_{14}$ is displayed for comparison (middle). Right panels: Masses around C$_{27}$H$_{14}$, C$_{28}$H$_{14}$, and C$_{29}$H$_{14}$. The corrected (real) molecular distributions are indicated by red error bars.} \label{fig_MALDIzoom}
\end{center}\end{figure*}

For the HPLC extract ``7 $-$ 35 min'', Fig. \ref{fig_MALDIzoom} presents expanded views of the mass spectrum around 400 amu (top left panel) and 350 amu (right panels). The peak patterns are representative for the mass ranges around other carbon numbers. The theoretical isotopic pattern of C$_{32}$H$_{14}$ is displayed in the center left panel.
The contributions of all higher-mass isotopologues (containing $^{13}$C and also $^{2}$H) were subtracted from the mass spectra to obtain the deconvoluted mass distributions, which are indicated by red error bars in the lower left panel as well as in the right panels of Fig. \ref{fig_MALDIzoom}. The strongest peak in the MALDI-TOF spectrum appears at a mass corresponding to a PAH with 28 C atoms. This agrees well with the strongest signal from benzocoronene (C$_{28}$H$_{14}$) that was observed in the HPLC spectrum \citep{steglich12}. The areas of the neighboring mass peaks caused by PAHs composed of 29 and 27 C atoms are comparatively weaker by 20 to 50 \%. Normal PAHs, like C$_{28}$H$_{14}$, or PAHs containing only one methylene group, like C$_{27}$H$_{14}$, produce stronger signals than those molecules with (many) excess hydrogens or methyl groups. The seemingly high aromaticity contradicts the results from the NMR measurements of the dissolved material, which are discussed in the next subsection, and it is also in contrast to the IR measurements presented later. This may be explained by an observational bias, as the intensity of the MALDI-TOF signals depends on the UV absorption behavior of the individual components of the PAH mixture. It can be expected that PAHs with strong absorption at 337 nm (the wavelength of nitrogen laser used in MALDI) experience a preferential ionization \citep{rader96}. Despite these drawbacks and a limited quantitative interpretability, we successfully applied the MALDI-TOF method to determine the maximum molecular sizes and the overall molecular size distributions of the fractionated PAH blends.

\subsection{NMR analysis} \label{sec_NMR}
\begin{figure*}\begin{center}
\epsscale{2.15} \plotone{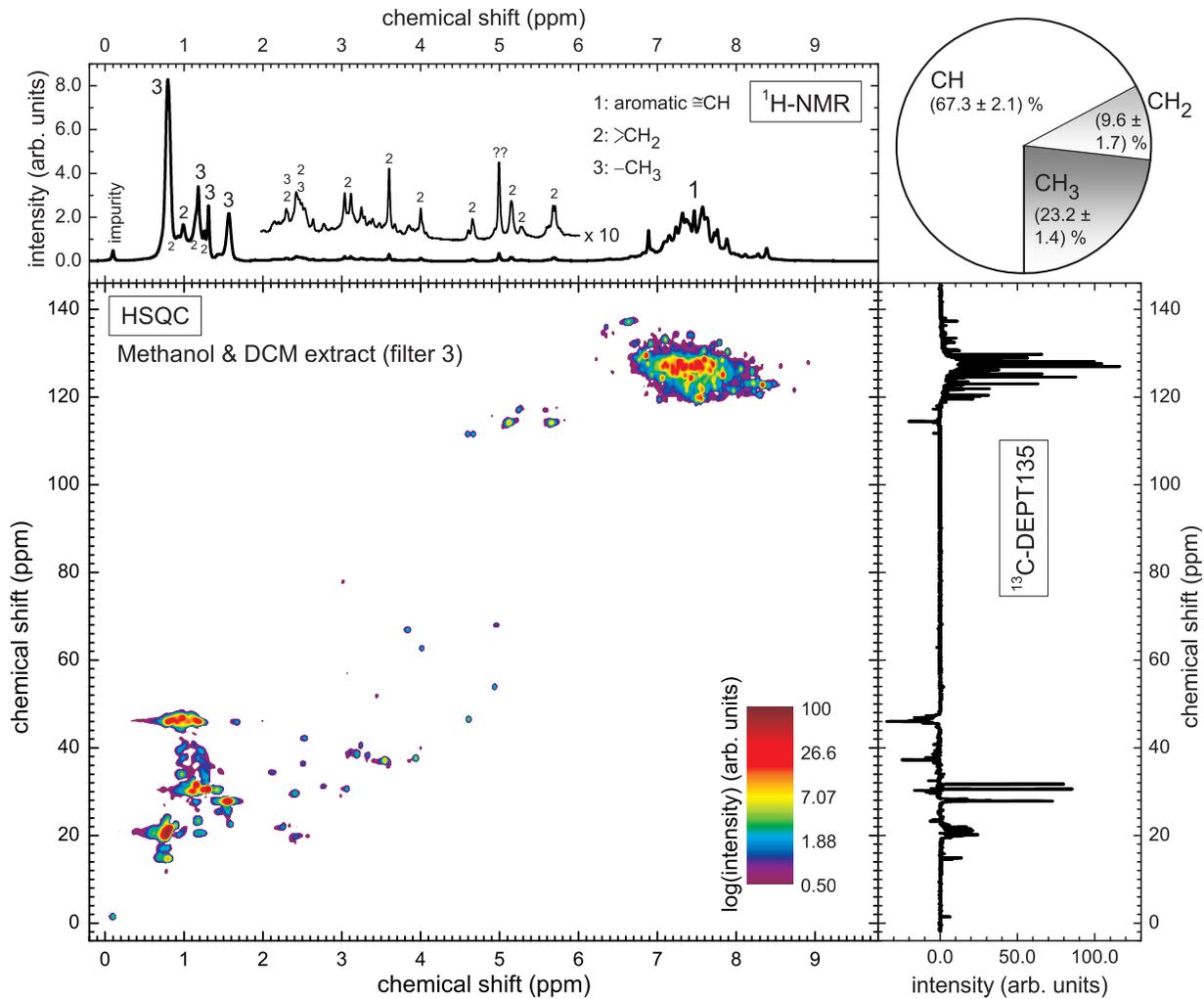} \caption{Complete NMR characterization of an extract from the laser pyrolysis (combined methanol and DCM extract). The amounts of \protect\includegraphics[scale=0.18]{arom-bond.eps}CH, \protect\includegraphics[scale=0.18]{doublebond.eps}CH$_2$, and \sbond CH$_3$ groups (upper right corner) were determined from the line areas of the $^1$H NMR experiment. The assignment of the different functional groups was carried out by connecting the peaks of the $^{13}$C DEPT135 scan with the peaks of the $^1$H scan via an HSQC experiment.} \label{fig_NMRexample}
\end{center}\end{figure*}

The NMR characterization of the various laser pyrolysis extracts was carried out on a Bruker Avance 600 MHz instrument. By combining different NMR experiments, we determined the relative abundances of \includegraphics[scale=0.18]{arom-bond.eps}CH, \includegraphics[scale=0.18]{doublebond.eps}CH$_2$, and \sbond CH$_3$ groups in our samples. Since the chemical shifts of aliphatic \includegraphics[scale=0.18]{doublebond.eps}CH$_2$ and \sbond CH$_3$ groups linked to polycyclic aromatic structures are very similar\footnote{A comprehensive collection of $^1$H and $^{13}$C NMR spectra of PAHs with various functional groups can be found in the catalogues edited by \citet{karcher85, karcher88, karcher91}.}, simple $^1$H and $^{13}$C NMR experiments were not sufficient to differentiate between both contributions. In addition, we performed measurements applying the distortionless enhancement by polarization transfer (DEPT) and heteronuclear single quantum coherence (HSQC) techniques. The complete analysis is depicted in Fig. \ref{fig_NMRexample}. The abundances of the different functional CH$_x$ groups can only be determined from the $^1$H NMR experiment as the integrated line intensities are directly proportional to the respective number of protons only in this spectrum. A discrimination between \includegraphics[scale=0.18]{doublebond.eps}CH$_2$ and \sbond CH$_3$ groups is possible with a $^{13}$C DEPT experiment. The $^{13}$C DEPT spectrum was measured at an angle of 135$^\circ$ (DEPT135). Positive peaks at chemical shifts larger than 110 ppm are due to aromatic \includegraphics[scale=0.18]{arom-bond.eps}CH groups, positive peaks at chemical shifts smaller than about 70 ppm are caused by \sbond CH$_3$ groups, and  negative peaks are from \includegraphics[scale=0.18]{doublebond.eps}CH$_2$ groups. Carbon atoms with no hydrogens attached produce no signal in a $^{13}$C DEPT scan. In contrast to a $^1$H NMR spectrum, the line intensities are not proportional to the number of nuclei. The HSQC experiment connects both spectra. Cross peaks in the two dimensional HSQC measurement link the assigned peaks of the $^{13}$C DEPT135 scan with the peaks of the $^1$H NMR scan, whose integrated intensities yield the abundance information.

For the different laser pyrolysis extracts, the relative abundances of \includegraphics[scale=0.18]{arom-bond.eps}CH, \includegraphics[scale=0.18]{doublebond.eps}CH$_2$, and \sbond CH$_3$ groups as well as the total hydrogen contents in aromatic and aliphatic groups are summarized in Table \ref{tab_4}. The values were determined by integrating the individual lines in the $^1$H NMR spectra. All peaks appearing at chemical shifts above 6 ppm were assigned to aromatic \includegraphics[scale=0.18]{arom-bond.eps}CH groups. Those below 6 ppm were attributed to \includegraphics[scale=0.18]{doublebond.eps}CH$_2$ and \sbond CH$_3$ groups. Between 0.5 and 3 ppm, the spectra feature many overlapping lines, which we tried to disentangle by deconvolution with Lorentzian line shapes. However, this procedure is not free of ambiguities. Accordingly, the uncertainties of the determined methylene and methyl abundances are somewhat larger.

Impurities from water might systematically contribute a small proportion to the overall hydrogen contents in aliphatic groups. A weak peak at 0.1 ppm present in all $^1$H NMR spectra is caused by vacuum grease and was excluded from the analysis. Further impurities, especially various small hydrocarbons, are expected in only negligible quantities. We found weak signatures of carbon-oxygen bonds in the $^{13}$C NMR spectra of three of our samples correlating well with the IR data. The $^{13}$C NMR spectrum of the HPLC extract 7-35 min (filter 1 \& 2) features a weak peak at 168 ppm, presumably caused by a \includegraphics[scale=0.18]{doublebond.eps}C\dbond O or \includegraphics[scale=0.18]{arom-bond.eps}C\sbond OH group. This sample also displays a quite strong CO stretching signal in the IR absorption spectrum. Very weak peaks between 168 and 173 ppm were also found for the DCM extract (filter 4) and the HPLC extract 7-20 min (filter 4). For the HPLC extract 20-35 min (filter 4), only the results from a $^1$H NMR experiment, i.e., the hydrogen content in aromatic or aliphatic groups, could be determined as the available amount of material was insufficient for a $^{13}$C NMR or a $^{13}$C DEPT135 experiment.

\begin{table*}
 \caption{Distributions of the H Atoms and CH$_x$ Groups in the Various Extracts}
\begin{tabular}{lccccc} \hline
 sample & \multicolumn{2}{c}{H content in} &  \multicolumn{3}{c}{No. of CH$_x$ groups}\\
 & aromatic CH & CH$_2$/CH$_3$ & $x$ = 1 & $x$ = 2 & $x$ = 3 \\ \hline \hline

 Meth. \& DCM$^a$ & \small 45.5 $\pm$ 1.4 &  \small 54.5 $\pm$ 1.9 &  \small 67.3 $\pm$ 2.1 &  \small 9.6 $\pm$ 1.7 &  \small 23.2 $\pm$ 1.4 \\

 DCM$^b$ & \small 11.5 $\pm$ 0.5 & \small 88.5 $\pm$ 2.7 & \small 25.3 $\pm$ 1.0 & \small 30.9 $\pm$ 4.9 & \small 43.9 $\pm$ 5.5 \\

 HPLC 7-20 min$^b$ & \small 4.4 $\pm$ 0.4 & \small 95.6 $\pm$ 2.9 & \small 10.6 $\pm$ 0.9 & \small 39.4 $\pm$ 5.7 & \small 50.1 $\pm$ 5.7 \\ 

 HPLC 7-35 min$^c$ & \small 8.6 $\pm$ 0.4 & \small 91.4 $\pm$ 2.8 & \small 18.3 $\pm$ 0.7 & \small 52.1 $\pm$ 6.5 & \small 29.5 $\pm$ 4.3 \\

 HPLC 20-35 min$^b$ & \small 8.5 $\pm$ 5.0 & \small 91.5 $\pm$ 5.0 & \small $--$ & \small $--$ & \small $--$ \\  \hline
\end{tabular}\\

\begin{tabular}{l} \footnotesize
\textbf{Note.} Distributions determined by NMR and expressed in terms of percentage. \\ \footnotesize
$^a$ filter 3, $^b$ filter 4, $^c$ filter 1 \& 2 \\
\end{tabular}
\label{tab_4}
 \end{table*}

The combined methanol and DCM extract is the least fractionated sample and it contains many small PAHs. The NMR analysis of this material reveals a high aromatic \includegraphics[scale=0.18]{arom-bond.eps}CH  abundance. The molecules in the other samples are characterized by a strong coverage with aliphatic hydrocarbon groups. While the DCM extract and the HPLC extract 7-20 min contain more \sbond CH$_3$ than \includegraphics[scale=0.18]{doublebond.eps}CH$_2$ groups, it is the other way around for the HPLC extract 7-35 min. These structural differences also lead to variations in the IR absorption spectra, enabling a detailed analysis of the 3.4 $\mu$m absorption complex and an unambiguous identification of its subpeaks.

We should point out that we do not exclude the possible presence of longer alkyl chains (\sbond CH$_2$\sbond CH$_3$, \sbond CH$_2$\sbond CH$_2$\sbond CH$_3$, ...) or vinyl groups (\sbond CH\dbond CH$_2$) connected to the PAHs of the laser pyrolysis samples. For mixtures containing many different species, it is very difficult, if not impossible, to differentiate between the methyl and methylene groups inside of those side chains and methyl or methylene groups directly attached to the hexagonal network. Most importantly, these two different structural configurations contribute in a similar or identical way not only to the NMR spectra, but also to the IR absorption spectra. While low concentrations of PAHs with ethyl groups (\sbond CH$_2$\sbond CH$_3$) were detected in toluene extracts from the laser pyrolysis \citep{jager06, jager07}, we have no direct evidence for the longer alkyl chains or vinyl groups. Furthermore, the presence of polyynyl-substituted PAHs can be excluded because the IR absorption spectra (Section \ref{sec_mixPAHs}) are void of acetylenic CH stretching bands expected at 3260 $-$ 3300 cm$^{-1}$ \citep{rouille12}. For the same reasoning, we rule out perceivable amounts of diamond-like \includegraphics[scale=0.18]{triplebond.eps}CH groups. Finally, structural exceptions from the perfect hexagonal network of ideal PAHs, such as pentagons and heptagons or cross-linked (biphenyl-like) PAHs can be present in some of the molecules of the pyrolysis samples as they have no specific impact on the overall spectroscopic properties. In Sections \ref{sec_indiv} and \ref{sec_theoryIR}, which are devoted to the determination of IR absorption cross sections of individual PAHs and mixtures of PAHs, we will restrict ourselves to those molecular structures that are most abundant in the laser pyrolysis samples, i. e., normal PAHs and PAHs with methyl and methylene groups attached directly to the hexagonal network.

\section{IR spectra of individual molecules} \label{sec_indiv}
\subsection{Experimental and theoretical methods}
The samples for the IR scans were obtained from commercial sources (Sigma-Aldrich, Campro Scientific) and used without further purification. We did not notice obvious impurity bands in our spectra. IR spectral measurements were performed on a Bruker 113v Fourier transform IR spectrometer at a resolution of 2 cm$^{-1}$. Every spectrum was measured for grains of the respective sample powder embedded in a CsI pellet at room temperature, prepared with a sample-to-matrix mass ratio of 1:500. An advantage of the pellet technique is the ability to obtain reliable information about the absorption cross sections of big molecules, which is useful for abundance estimates and theory calibrations. The spectra thus obtained reflect the absorption behavior of PAHs as constituents of carbonaceous grains in astrophysical environments. Effects from temperature differences can be almost neglected in this case. Only minor blue shifts of bands amounting to less than 1 \% of the vibrational energy are to be expected upon temperature decrease.

For comparisons and for later abundance estimates, DFT calculations were carried out for each molecule. The computations were performed with the Gaussian09 software package \citep{frisch09} applying the B3LYP functional \citep{becke88, becke93, lee88} in conjunction with the 6-31G basis set \citep{hehre72}. Using this method, we optimized the molecular geometries and, subsequently, we calculated the vibrational modes for every molecule. The absence of imaginary frequencies in the theoretical results ensured that the optimized geometries corresponded to the minima of the potential energy surfaces. Finally, the theoretical IR spectra were obtained by convoluting the harmonic frequencies and intensities of the different vibrational modes with Lorentzian functions with full width at half maximum of 5 cm$^{-1}$.  We also tested larger basis sets as well as more advanced functionals. The only notable differences, however, were the scaling factors that had to be applied in order to approach the laboratory spectra.

To bring the DFT-calculated frequencies of PAH vibrations into agreement with laboratory data, the theoretical values had to be scaled by a factor smaller than one. Usually, two different scaling factors are applied, one for the CH stretching modes and one for all other modes. We follow this route here and, instead of determining our own values, we simply apply the same frequency scaling factors that were already obtained by \citet{bauschlicher97} for normal PAHs. At the same level of theory, similar values of the scaling factors were found for PAHs carrying vinyl side groups \citep{maurya12}. For our purpose, this approach also works sufficiently well for vibrations involving aliphatic groups in polycyclic aromatic compounds.

While the calculated intensities of non-CH stretching modes of PAHs in the solid phase display an acceptable agreement with quantitative laboratory spectra (see results hereafter), the absorption strengths of the CH stretching modes are about two to four times over-estimated. The discrepancy is likely due to the embedding of the molecules inside grains. \citet{joblin94} have shown that, compared to gas phase molecules, the aromatic CH stretching bands of PAHs in the solid phase are suppressed by a factor of two to four. To what extent the oscillator strengths of aliphatic stretching vibrations in the gas phase differ from the solid phase is unknown and beyond the scope of this investigation. We scaled the computed intensities of all CH stretching vibrations such that the theoretical cross sections are identical to the experimental ones. The intensity scaling factors for the aromatic \includegraphics[scale=0.18]{arom-bond.eps}CH, the aliphatic \includegraphics[scale=0.18]{doublebond.eps}CH$_2$, and the aliphatic \sbond CH$_3$ groups were determined separately. All results concerning absolute values are applicable to molecules in grains only. The absorption cross sections of the experimental spectra were obtained by either integration or band fitting. The first method was used in the case of the hydrogenated PAHs for which the vibrational modes involving the \includegraphics[scale=0.18]{arom-bond.eps}CH and \includegraphics[scale=0.18]{doublebond.eps}CH$_2$ groups had a clear energetic separation, while the fitting procedure was applied to the PAHs with methyl groups displaying overlapping bands from \includegraphics[scale=0.18]{arom-bond.eps}CH and \sbond CH$_3$ groups. All (average) scaling factors are summarized in Table \ref{tab_1}. For the intensity scaling factor of aromatic CH stretching vibrations, additional laboratory data from the molecules Ant (C$_{14}$H$_{10}$) and Phn (C$_{14}$H$_{10}$) were used (not shown here).

\begin{table*}
 \caption{Scaling Factors Applied to the Various Vibrational Modes}
\begin{tabular}{lcc} \hline
  mode & intensity scaling & frequency scaling \\ \hline \hline
  aromatic CH stretching & $0.27 \pm 0.06$$^a$ & 0.9552$^d$ \\
  methylenic CH stretching & $0.41 \pm 0.12$$^b$ & 0.9552$^d$ \\
  methylic CH stretching & $0.45 \pm 0.14$$^c$ & 0.9552$^d$ \\
  all other modes & $1.0$ & 0.9610$^d$ \\ \hline
\end{tabular}\\
\begin{tabular}{l} \footnotesize
$^a$ Determined from Ant, Phn, and from the same molecules as in Figs. \ref{fig_DiH-PAHs} and \ref{fig_CH3-PAHs}. \\ \footnotesize
$^b$ Determined from the same molecules as in Fig. \ref{fig_DiH-PAHs}. \\ \footnotesize
$^c$ Determined from the same molecules as in Fig. \ref{fig_CH3-PAHs}. \\ \footnotesize
$^d$ Taken from \citet{bauschlicher97}. \\
\end{tabular}
\label{tab_1}
 \end{table*}

\subsection{Hydrogenation of peripheral C atoms} \label{sec_outer}
Astrophysically relevant IR measurements of a few individual and rather small PAHs with excess hydrogen on the outer C atoms (\includegraphics[scale=0.18]{doublebond.eps}CH$_2$) were already presented by \citet{beegle01}, \citet{bernstein96}, \citet{fu12}, \citet{ricks09}, \citet{szczepanski11}, and \citet{wagner00}. DFT calculations of IR spectra were performed by \citet{beegle01}, \citet{pauzat01}, and \citet{szczepanski11}. In this subsection, we will consider only such compounds that show excess hydrogenation on the periphery. Peripheral C atoms are bonded to only two other C atoms. \citet{may00} and \citet{rauls08} have shown that due to lower energy barriers for hydrogen addition and considerably higher dissociation energies the formation of sp$^3$ centers is more likely to occur on outer C atoms. The inner part of a larger PAH remains aromatic as long as the periphery is not fully saturated.

\begin{figure*}\begin{center}
\epsscale{2.15} \plotone{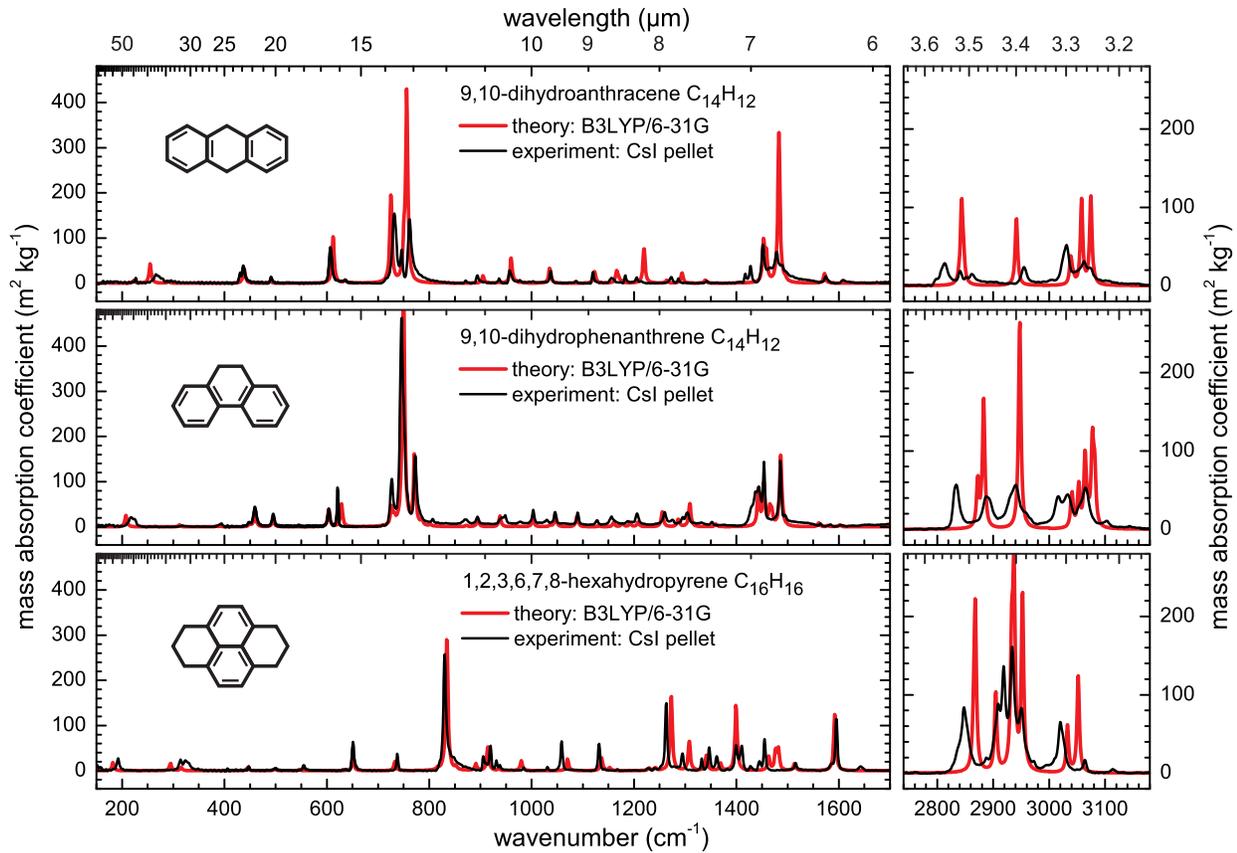} \caption{Measured spectra of grains of partly hydrogenated PAHs embedded in CsI pellets in comparison to scaled theoretical spectra calculated at the B3LYP/6-31G level of theory.} \label{fig_DiH-PAHs}
\end{center}\end{figure*}

The three selected partly hydrogenated PAHs discussed in this subsection are 9,10-dihydroanthracene (C$_{14}$H$_{12}$; H2Ant), 9,10-dihydrophenanthrene (C$_{14}$H$_{12}$; H2Phn), and 1,2,3,6,7,8-hexahydropyrene (C$_{16}$H$_{16}$; H6Pyr). They represent three different kinds of \includegraphics[scale=0.18]{doublebond.eps}CH$_2$ groups, namely solo- (H2Ant), duo- (H2Phn), and trio-CH$_2$ (H6Pyr), where in the last case three \includegraphics[scale=0.18]{doublebond.eps}CH$_2$ groups are placed next to each other. The measured IR spectra, obtained for grains of the sample powder embedded in CsI pellets, along with the scaled theoretical spectra are presented in Fig. \ref{fig_DiH-PAHs}. For each molecule, we applied separate scaling factors to the band intensities of the stretching vibrations, chosen such that the calculated cross sections match the experimental ones.

Except for the absolute intensities of two vibrational modes of H2Ant at 756 and 1483 cm$^{-1}$, which are somewhat overestimated by the calculation, and slightly asymmetric band profiles present in all spectra, which are an artifact of the pellet measurements, the spectra in Fig. \ref{fig_DiH-PAHs} display a good agreement between theory and experiment for frequencies smaller than 1700 cm$^{-1}$. The laboratory data above 2800 cm$^{-1}$, however, is less well reproduced by the calculations. The discrepancies can, for example, arise from couplings between the first overtones of the CH$_2$ deformation modes seen between 1400 and 1500 cm$^{-1}$ and the stretching modes between 2800 and 3000 cm$^{-1}$. Such mode couplings are not accounted for by the harmonic approximation made in the frequency calculations.

\subsection{Hydrogenation of inner C atoms} \label{sec_HinnerC}
\begin{figure*}\begin{center}
\epsscale{2.15} \plotone{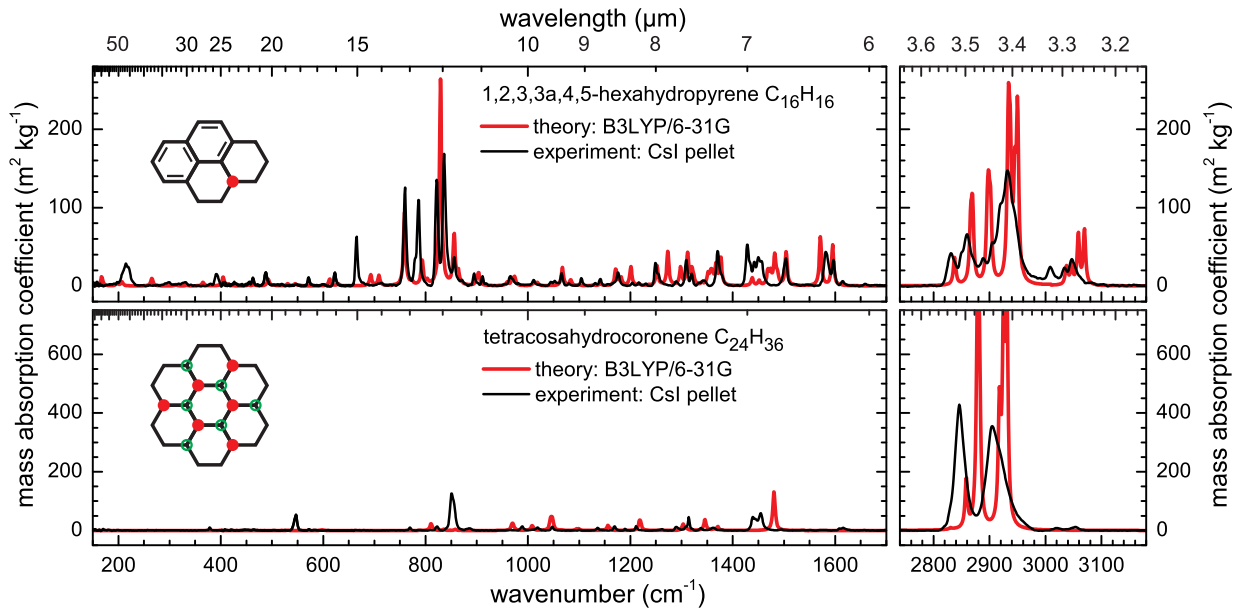} \caption{Measured spectra of grains of PAHs with excess hydrogenation on outer and inner C atoms embedded in CsI pellets compared to scaled theoretical spectra calculated at the B3LYP/6-31G level of theory. Only the displayed isomers were calculated. The empty green and filled red circles on the molecular structures mark H atoms situated above or below the plane of C atoms.} \label{fig_other_H-PAHs}
\end{center}\end{figure*}

As mentioned before, the dissociation energies for hydrogen removal from inner C atoms are considerably lower compared to outer C atoms. Nevertheless, in regions of space with low photon fluxes, PAHs may reach a strong hydrogen coverage, possibly leading to superhydrogenated structures \citep[e.g.,][]{rauls08}. In Fig. \ref{fig_other_H-PAHs}, we compare the spectra of two PAHs with H atoms attached to outer and, additionally, to inner C atoms. The first molecule, 1,2,3,3a,4,5-hexahydropyrene (C$_{16}$H$_{16}$), contains only one tertiary \includegraphics[scale=0.18]{triplebond.eps}CH group, alongside with five \includegraphics[scale=0.18]{doublebond.eps}CH$_2$ and five aromatic \includegraphics[scale=0.18]{arom-bond.eps}CH groups. In contrast, the second molecule, tetracosahydrocoronene (C$_{24}$H$_{36}$), is completely hydrogenated. Altogether, it contains twelve \includegraphics[scale=0.18]{doublebond.eps}CH$_2$ and twelve \includegraphics[scale=0.18]{triplebond.eps}CH groups, which use up all $\pi$ electrons of the parent aromatic structure. In these molecules, the H atoms can be situated above or below the aromatic plane, leading to different structural isomers. No information about the isomeric distributions of the samples was available. The scaled theoretical spectra were computed only for those molecular structures that are displayed in Fig. \ref{fig_other_H-PAHs}, whereby the empty green and filled red circles mark H atoms placed above and below the plane of C atoms, respectively. The neglect of other possible isomers likely explains the discrepancies between the laboratory and theoretical data. It should be mentioned that the CH stretching region of the spectrum of C$_{24}$H$_{36}$ isolated in solid Ar was already presented by \citet{bernstein96}.

In contrast to spectra of only peripherally hydrogenated PAHs, the low-energy part of the 3.4 $\mu$m absorption complex at 3.51 $\mu$m (2846 cm$^{-1}$) is exceptionally strong in C$_{24}$H$_{36}$. The 3.51 $\mu$m band arises from stretching vibrations of H atoms, which are situated above or below the plane of the carbon framework (``out-of-plane'' hydrogens). Naturally, it is intense in molecules containing many tertiary \includegraphics[scale=0.18]{triplebond.eps}CH groups.

\subsection{Methyl groups on outer C atoms}
Methylated PAHs were already discussed in an astrophysical context by a few authors. Based on an absorption spectrum of hot methylcoronene in the gas phase, showing the CH stretching region of the molecule, \citet{joblin96} suggested that these species could contribute to the 3.4 $\mu$m features in astrophysical emission sources. However, this assignment was later abandoned when partly hydrogenated PAHs produced a better match in terms of band positions \citep[e.g.,][]{bernstein96, wagner00}. \citet{wagner00} also presented the emission spectra of the CH stretching vibrations of 2-methylnaphthalene and 2-methylphenanthrene. The absorption spectra of several gas phase dimethylnaphthalenes can be found in the publication of \citet{das08}. Additional laboratory data of a few PAHs with methyl groups in pellets or in solution are, for instance, available in the spectral atlas edited by \citet{karcher91}. Results from DFT calculations were presented by \citet{bauschlicher98} and calculated IR spectra are also accessible online (http://www.astrochem.org/pahdb).

\begin{figure*}\begin{center}
\epsscale{2.15} \plotone{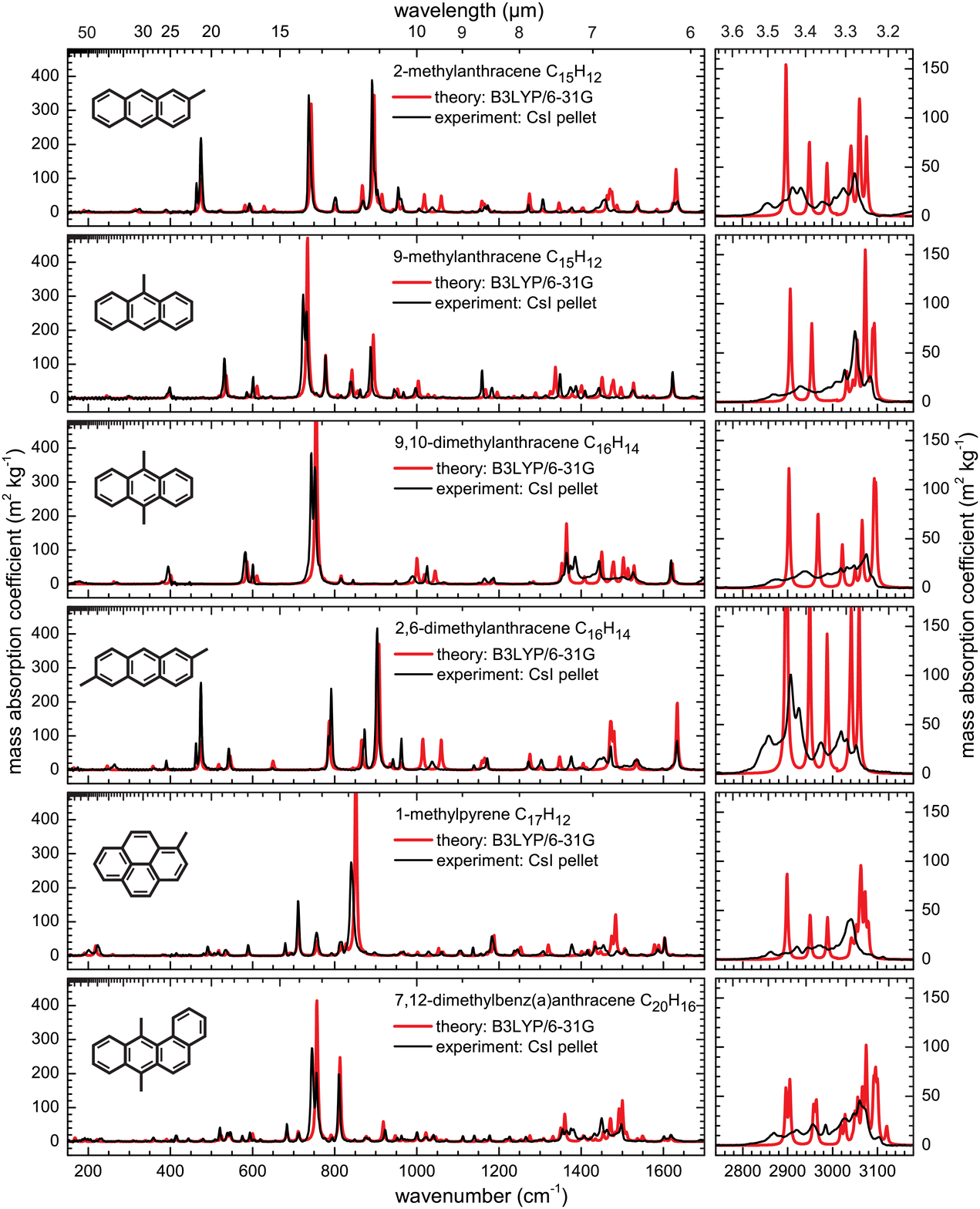} \caption{Measured spectra of grains of methylated PAHs embedded in CsI pellets in comparison with scaled theoretical spectra calculated at the B3LYP/6-31G level of theory.} \label{fig_CH3-PAHs}
\end{center}\end{figure*}

In Fig. \ref{fig_CH3-PAHs}, the absorption spectra of methylated PAHs embedded as grains in CsI pellets are displayed. The theoretical spectra calculated at the B3LYP/6-31G level of theory are shown for comparison. The average intensity scaling factor of stretching vibrations in \sbond CH$_3$ groups is given in Table \ref{tab_1}. As in the case of hydrogenated PAHs, mode couplings probably disturb the aliphatic CH stretching bands in PAHs with methyl groups and cause some discrepancies between the experimental and calculated spectra (i.e., there is a possible interaction with the first overtones of the CH$_3$ bending modes seen between 1400 and 1500 cm$^{-1}$).

\subsection{Average spectra}
\begin{figure*}\begin{center}
\epsscale{2.15} \plotone{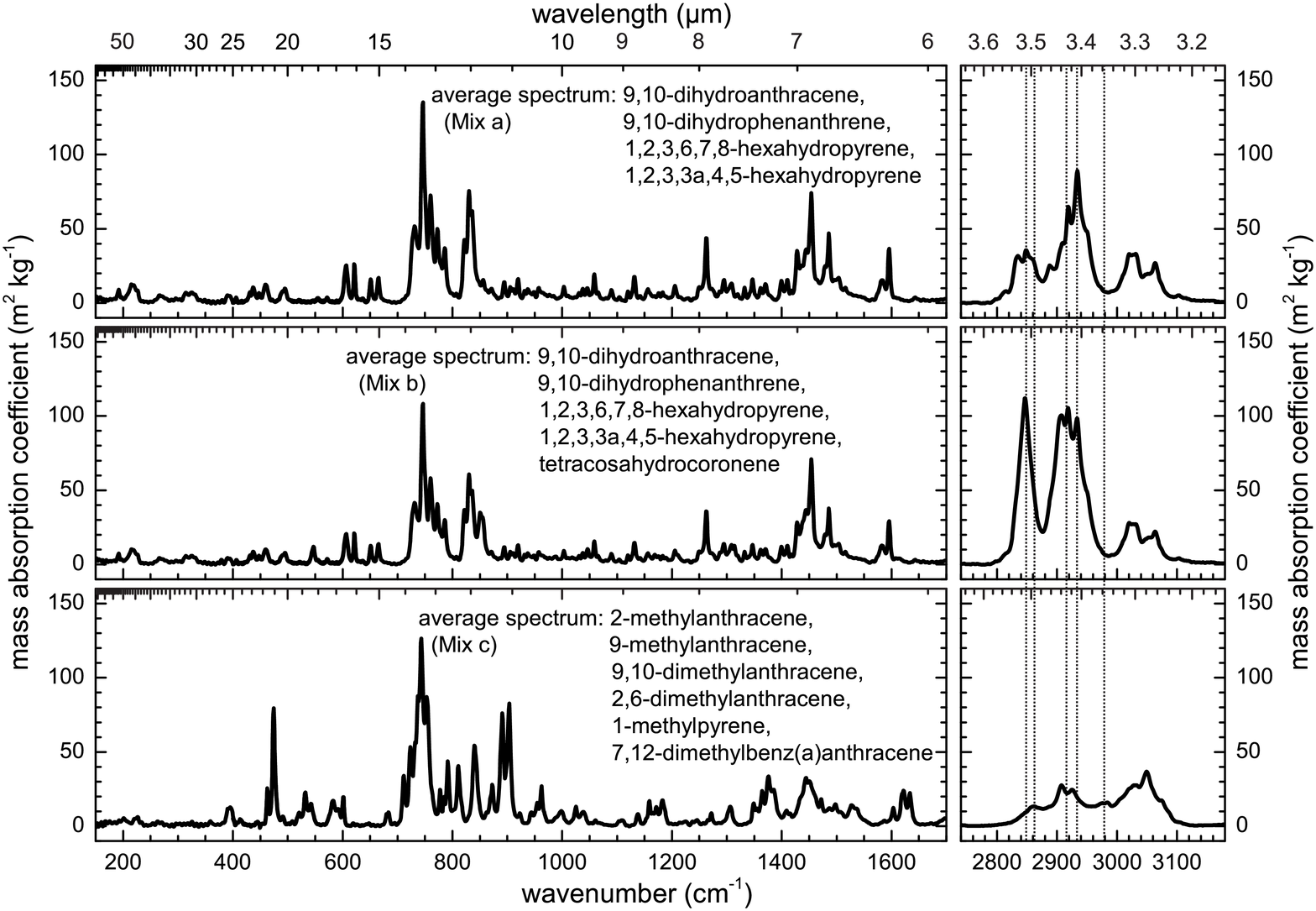} \caption{Average spectra of the pellet measurements presented in Figs. \ref{fig_DiH-PAHs}, \ref{fig_other_H-PAHs}, and \ref{fig_CH3-PAHs}. Vertical dotted lines mark prominent peaks in the aliphatic CH stretching region and serve as guide to the eye.} \label{fig_average_spectra}
\end{center}\end{figure*}

In Fig. \ref{fig_average_spectra}, three spectra are displayed, which were obtained by numerically averaging the previously presented spectra of individual molecules. The first mixture (Mix a) contains the moderately hydrogenated PAHs H2Ant, H2Phn, and two different hexahydropyrenes. In the second mixture (Mix b), the completely hydrogenated molecule tetracosahydrocoronene was added. The last mixture is composed of all six methylated PAHs from Fig. \ref{fig_CH3-PAHs}. If sufficiently different species contribute to a molecular blend then features characteristic to certain individual structures, like the appearance or non-appearance of bands for symmetry reasons, smooth out and only those features remain, which are representative for the functional groups defining the class of investigated molecules. Even though rather few spectra were averaged in our case, this tendency can already be observed in Fig. \ref{fig_average_spectra}. Aromatic \includegraphics[scale=0.18]{arom-bond.eps}CH groups give rise to blends of bands from stretching vibrations around 3040 cm$^{-1}$ (3.3 $\mu$m) as well as out-of-plane bending vibrations at about 840 cm$^{-1}$ (11.9 $\mu$m; duo CH) and 750 cm$^{-1}$ (13.3 $\mu$m; trio/quartet CH). Aliphatic features from bending or deformation modes in \includegraphics[scale=0.18]{doublebond.eps}CH$_2$ and \sbond CH$_3$ groups appear at 1263 cm$^{-1}$ (7.9 $\mu$m; CH$_2$), 1375 cm$^{-1}$ (7.3 $\mu$m; CH$_3$), and 1450 cm$^{-1}$ (6.9 $\mu$m; CH$_2$ and CH$_3$). Bands from CC stretching vibrations show up around 1600 cm$^{-1}$ (6.3 $\mu$m). Of particular interest for the comparison with observational data from the diffuse ISM is the spectral region of aliphatic CH stretching vibrations around 3.4 $\mu$m. The PAHs with excess hydrogenation (Mix a, b) exhibit two band clusters at 2847 cm$^{-1}$ (3.51 $\mu$m) and at about 2920 cm$^{-1}$ (3.42 $\mu$m). The band cluster at lower energy is considerably enhanced when molecules with hydrogenation on inner C atoms, i.e., \includegraphics[scale=0.18]{triplebond.eps}CH groups, are included. Theory suggests that it is caused by stretching motions involving ``out-of-plane'' hydrogens. In PAHs with \sbond CH$_3$ groups, three distinct bands emerge at 2863 cm$^{-1}$ (3.49 $\mu$m), 2915 cm$^{-1}$ (3.43 $\mu$m), and 2980 cm$^{-1}$ (3.36 $\mu$m). We discuss their theoretical interpretation in more detail in the following section.

\section{Theoretical IR spectra of molecular mixtures} \label{sec_theoryIR}
\subsection{The composition of the mixtures}
In this Section, we will use the previously obtained scaling factors for DFT calculations at the B3LYP/6-31G level of theory (Table \ref{tab_1}) to compute the artificial IR spectra of molecular mixtures composed of a larger set of hydrogenated and methylated PAHs. This approach will help us to better understand the vibrational modes involved and to disentangle the contributions from \includegraphics[scale=0.18]{arom-bond.eps}CH, \includegraphics[scale=0.18]{doublebond.eps}CH$_2$, and \sbond CH$_3$ groups, which are present in the complicated blends of molecules from the laser pyrolysis. It also provides absolute values for the absorption cross sections expected for grains composed of PAH mixtures. Of particular interest for comparison with astrophysical data in the CH stretching wavelength region will be the inferred absorption strength per CH$_x$ group ($x$ = 1, 2, 3).

\begin{figure*}\begin{center}
\epsscale{2.15} \plotone{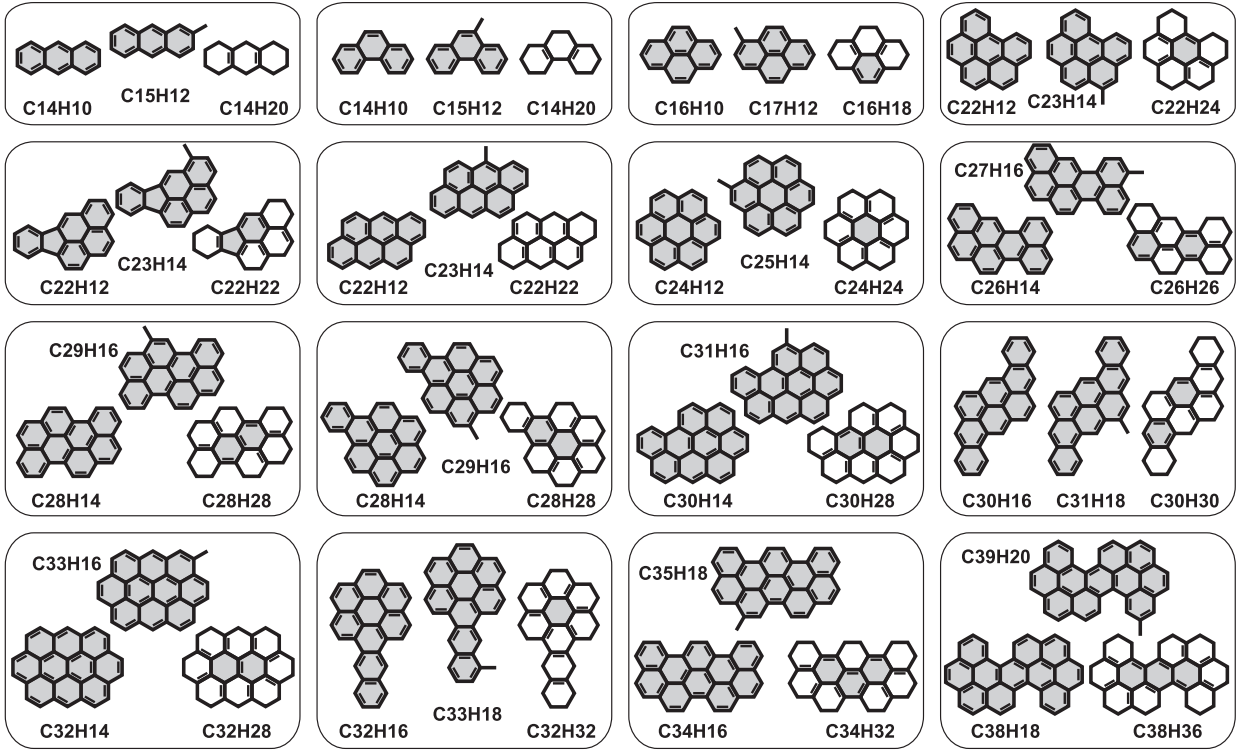} \caption{Overview over the normal, methylated, and hydrogenated PAHs, which are included in the calculations shown in Fig. \ref{fig_H-PAHs_calc}. Fully aromatic carbon rings are marked in gray.} \label{fig_overviewPAHs}
\end{center}\end{figure*}

We calculated the spectra for three different sets of molecules. The first set, representing the fully aromatic compounds, is composed of those PAHs that were previously identified by HPLC in the extracts from the laser pyrolysis \citep[see Fig. 3 in][]{steglich12}. Here, we only chose those molecules that contain more than 22 C atoms and, in addition, the most frequent small PAHs Ant, Phn, and Pyr, which are present as impurities in the high-mass extracts. The second set gathers methyl derivatives of the PAHs that form the first set. They were obtained by substituting an arbitrarily chosen H atom with a methyl group. Multiple substitutions were not considered to avoid an unnecessary increase of the molecular sizes (by adding C atoms), which would complicate a comparison with the first set. Finally, the third set consists of species formed by having the periphery of the molecules of the first set fully hydrogenated. As we consider only neutral species with a closed-shell electronic structure, it was in some cases necessary to keep two \includegraphics[scale=0.18]{arom-bond.eps}CH groups. All compounds included in the calculations are displayed in Fig. \ref{fig_overviewPAHs}. Fully aromatic hexagonal rings are marked in gray. We will continue to use the term ``hydrogenated PAHs'' throughout the text, even though some of the calculated molecules do not contain aromatic cycles anymore.

\subsection{DFT calculated spectra}
\begin{figure*}\begin{center}
\epsscale{2.15} \plotone{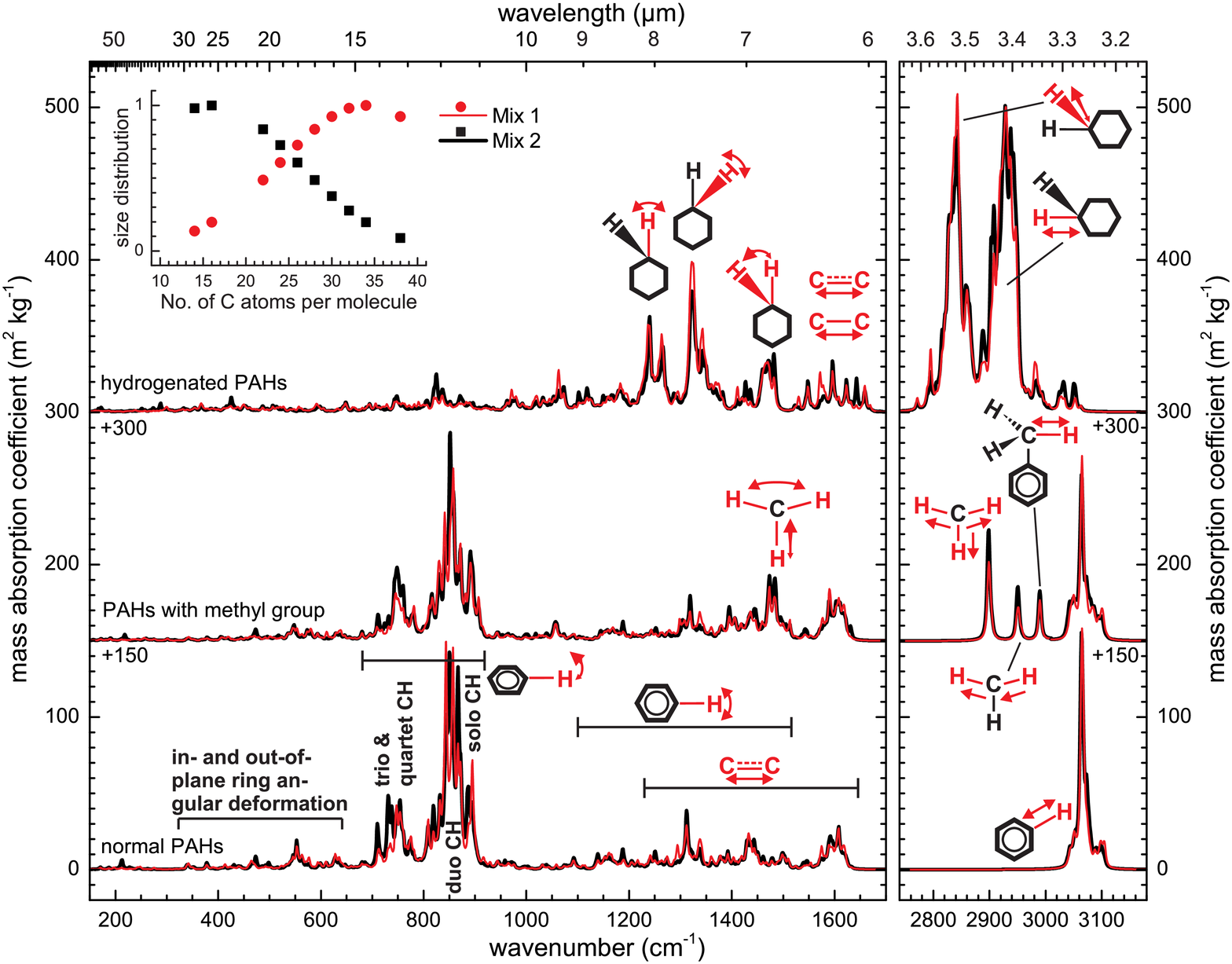} \caption{Theoretical absorption spectra as expected for grains composed of mixtures of normal (bottom), methylated (center), and peripherally hydrogenated PAHs (top). The vibrational modes are depicted in a simplified manner. Inset: Size distributions of the six different mixtures. In the case of the PAHs with methyl groups, each molecule contains one more C atom.} \label{fig_H-PAHs_calc}
\end{center}\end{figure*}

The theoretical spectra of the blends of normal, methylated, and fully hydrogenated PAHs are displayed in Fig. \ref{fig_H-PAHs_calc}. Two different size distributions, as depicted in the inset, were computed for every of the three blends. (In the case of the PAHs with \sbond CH$_3$ groups, each molecule actually contains one more C atom than pictured by the symbols.) The resulting two spectra for each blend are almost identical, featuring only minor variations among the CH$_x$ bending modes and basically overlapping each other in the CH$_x$ stretching region. In contrast to the electronic spectra \citep[see][]{steglich10}, the IR spectra are not very size-dependent as long as the relative amounts of certain functional groups (e.g., solo, duo, and trio \includegraphics[scale=0.18]{arom-bond.eps}CH groups) remain unaltered. Simplified pictures of the vibrations, which mainly contribute to the different blends of absorption bands, are also shown in Fig. \ref{fig_H-PAHs_calc}. The intensity scale we chose for display is in units of ``mass absorption coefficient'' (m$^2$ kg$^{-1}$). However, it might be useful to convert it into ``absorption cross section per C atom'' or ``absorption cross section per CH$_x$ group''. In that case, the conversion factors can be taken from Table \ref{tab_2}.

\begin{table*}
\caption{Conversion Factors for the Intensity Scale in Fig. \ref{fig_H-PAHs_calc}}
\begin{tabular}{lccc} \hline
                & normal PAHs & with \sbond CH$_3$ group & with \includegraphics[scale=0.18]{doublebond.eps}CH$_2$ groups \\ \hline\hline

mass abs. coefficient (m$^2$ kg$^{-1}$) & 1 & 1 & 1 \\
&&&\\

abs. cross section per C atom (10$^{-24}$ m$^{2}$) & 0.021 & 0.021 & 0.022 \\ 
&&&\\

      abs. cross section                            & $x$ = 1 & $x$ = 3 & $x$ = 2 \\
 per CH$_x$ group (10$^{-24}$ m$^{2}$) & 0.040 $\pm$ 0.002  & 0.492 (mix 2)  & 0.044 $\pm$ 0.001 \\
                                        &   & ... 0.635 (mix 1)$^a$ &   \\ \hline
\end{tabular}\\
\begin{tabular}{l} \footnotesize
$^a$ To be applied to stretching vibrations of CH$_3$ groups between 2890 and 3010 cm$^{-1}$ only. \\
\end{tabular}
\label{tab_2}
 \end{table*}

The computed spectra of the fully aromatic PAHs are dominated by CH stretching modes around 3065 cm$^{-1}$ and CH out-of-plane bending modes between 700 and 900 cm$^{-1}$.  Weaker bands arise from ring deformation (350 $-$ 650 cm$^{-1}$), CH in-plane bending (1100 $-$ 1500 cm$^{-1}$), and CC stretching vibrations (1250 $-$ 1650 cm$^{-1}$). The methyl groups in our second molecular set basically add four obvious bands to the spectra. These are caused by the deformation or bending modes at about 1480 cm$^{-1}$ and three different stretching modes of the \sbond CH$_3$ group between 2890 and 3000 cm$^{-1}$. The stretching mode at lowest energy is caused by a simultaneous, symmetric movement of all three H atoms. The other two modes are due to asymmetric motions, whereby the symmetry notion should be understood with respect to the methyl group only and should not be confused with the symmetry regarding the whole molecule. A methyl group attached to a PAH tends to be oriented in such a way that only one out of the three H atoms is in the same plane as the neighboring carbon ring(s). The other two hydrogens are situated above and below that plane, respectively. While the stretching motion of the ``in-plane'' hydrogen is responsible for the band at highest energy (2990 cm$^{-1}$), the asymmetric vibrations of both ``out-of-plane'' hydrogens give rise to the band in the center. The fully hydrogenated compounds (blend 3) show very strong IR activity in the frequency range 2800 $-$ 3000 cm$^{-1}$ due to aliphatic CH stretching vibrations and moderately strong IR activity in the range 1200 $-$ 1500 cm$^{-1}$ caused by deformation modes of CH$_2$ groups, accompanied by comparatively weak CC stretching vibrations between 1520 and 1670 cm$^{-1}$. The two H atoms of the \includegraphics[scale=0.18]{doublebond.eps}CH$_2$ group tend to be placed ``in-'' and ``out-of-plane'', giving rise to energetically separated deformation and stretching bands. The bending and stretching motions of the ``in-plane'' hydrogens produce bands around 1250 and 2930 cm$^{-1}$. The corresponding vibrations of the ``out-of-plane'' hydrogens appear at about 1330 and 2840 cm$^{-1}$. An additional band at 1470 cm$^{-1}$ is due to a scissor vibration of both H atoms.

We want to add a short remark here regarding the classical labeling of the CH stretching vibrations in carbonaceous compounds. The terms ``symmetric'' and ``asymmetric'' vibrations in \includegraphics[scale=0.18]{doublebond.eps}CH$_2$ groups, commonly used to describe the peaks in the 3.4 $\mu$m absorption complex, may give a wrong image about the vibrational modes in hydrogenated PAHs. First, it should be realized that, in the aforementioned context, the symmetry refers to the side group only. From the theoretical point of view, the two band clusters that are expected to characterize the 3.4 $\mu$m absorption own their energetic separation from the tendency of the H atoms being situated in or out of the plane of neighboring carbon rings. Usually, there are no clear symmetric or asymmetric modes of vibration confined to the \includegraphics[scale=0.18]{doublebond.eps}CH$_2$ group alone. However, there can be symmetric and asymmetric modes if the whole molecule is considered. Density functional theory does not predict strong energetic differences between these symmetric and asymmetric modes. Nevertheless, a distinct splitting can occur in the experimental spectra if Fermi resonances play a role (see Section \ref{sec_outer}). This is an effect specific for certain molecular structures, and it can loose its importance when averaging the spectra of sufficiently different molecules.

The reader may compare the calculated spectra of the molecular blends with the average experimental spectra from Fig. \ref{fig_average_spectra}. Note, however, that the calculated band positions, especially in the CH stretching region, might not perfectly coincide with the corresponding measured values as we used the same frequency scaling factor for aromatic and aliphatic vibrations. Although the measured and calculated blends are composed of different molecules, the common features for the methylated and hydrogenated sets are obvious. The computed spectral features of the methylated PAHs, for instance, agree well, both in terms of band positions and absolute intensities, with bands seen in the laboratory average spectrum (Mix c). The differences for the CH out-of-plane bending modes (700 $-$ 900 cm$^{-1}$) are due to different amounts of solo-, duo-, and trio-hydrogens. The most obvious discrepancies appear for the CH stretching modes because of Fermi resonances (see earlier discussion) and a stronger band broadening active in the experimental spectrum. Nevertheless, in agreement with theory, three band clusters of the stretching vibrations in the methyl group around 2860, 2915, and 2980 cm$^{-1}$ are recognizable in the laboratory spectrum. The hydrogenated PAHs, on the other hand, exhibit two band clusters from aliphatic CH stretching modes, both in theory and laboratory (Mix a, b). (Remember that, in contrast to the computed mixture of hydrogenated PAHs, the average experimental spectra display stronger aromatic features, because the molecules in that case are not fully hydrogenated on their periphery.) The most notable difference originates from the ``out-of-plane'' H atoms of the \includegraphics[scale=0.18]{doublebond.eps}CH$_2$ groups. The intensities of their bending and stretching vibrational modes calculated around 1330 and 2840 cm$^{-1}$ are somewhat overestimated by theory. Only when the totally hydrogenated molecule C$_{24}$H$_{36}$ is included (Mix b), does the laboratory low-energy CH stretching band reach an intensity comparable to the high-energy part of the aliphatic 3.4 $\mu$m feature.

\subsection{Absorption cross sections} \label{sec_cross-sect}
The integrated absorption cross sections of the aromatic and aliphatic stretching bands at 3.3 $\mu$m and around 3.4 $\mu$m can be used for the estimation of abundances of \includegraphics[scale=0.18]{arom-bond.eps}CH, \includegraphics[scale=0.18]{doublebond.eps}CH$_2$, and \sbond CH$_3$ functional groups in the ISM. We determined these values from the calculated spectra (Fig. \ref{fig_H-PAHs_calc}). They are summarized in Table \ref{tab_3}. The relatively large errors are due to the uncertainties of the applied intensity scaling factors. It is found that the integrated absorption strengths of \includegraphics[scale=0.18]{doublebond.eps}CH$_2$ and \sbond CH$_3$ groups in hydrogenated and methylated PAHs are almost identical and about seven times stronger than the absorption strength of an aromatic \includegraphics[scale=0.18]{arom-bond.eps}CH group. As mentioned before, these values are valid for PAHs in grains only. For gas phase molecules, the unscaled theoretical cross sections might be closer to reality \citep[see Table \ref{tab_1} and][]{joblin94}.

\begin{table}
 \caption{Integrated Absorption Cross Sections for the CH Stretching Modes in PAH Grains}
\begin{tabular}{ll} \hline
 functional group & cross section per CH$_x$ \\
 & (10$^{-20}$ cm) \\ \hline\hline
\includegraphics[scale=0.18]{arom-bond.eps}CH 3.3 $\mu$m & 98 $\pm$ 22 \\ 
& \\
\includegraphics[scale=0.18]{doublebond.eps}CH$_2$ 3.4 $\mu$m & 677 $\pm$ 198 \\
& \\
\sbond CH$_3$ 3.4 $\mu$m & 661 $\pm$ 206 \\ \hline
\end{tabular}\\
\label{tab_3}
 \end{table}

\section{Measured IR spectra of PAH blends with mixed aromatic/aliphatic character} \label{sec_mixPAHs}
\subsection{Experimental methods} \label{sec_exp}
To prepare the material for the spectral measurements we used two different methods. The first method involved embedding grains of the samples in CsI or KBr pellets with a fixed sample-to-matrix mass ratio of 1:500. The pellet technique allowed quantitative measurements of the same substance that was analyzed previously by NMR. The second preparation procedure was the matrix isolation technique employing solid Ne at cryogenic temperatures (6 K) as host material. Each sample was transferred via DCM solvent onto a CaF$_2$ substrate and placed into a high-vacuum chamber ($p$ $<$ 2 $\times$ 10$^{-6}$ mbar). The 532 nm output of a Nd:YAG laser evaporated the material and the molecular flow, together with an excess of Ne atoms, was directed toward a KBr window used for spectroscopy and cooled down by a closed-cycle cryostat. In all experiments, the laser was operated at 10 Hz, $\approx$ 5 ns pulse duration, and 16 mJ pulse energy. The pulse intensity on the sample surface was about 2 $\times$ 10$^5$ W/mm$^2$, which was close to the minimum power necessary to transfer the material into the gas phase. Compared to measurements using pellets, the matrix isolation spectroscopy (MIS) has the advantage that the individual molecules of the extract are isolated from each other, minimizing the spectral perturbations caused by intermolecular interactions and, thus, mimicking the conditions of free-flying molecules in the ISM. However, the downside is a gradual graphitization of the sample material's surface upon laser irradiation, which shields the molecules in the lower layers from further desorption. The formation of graphitic surface layers and, in general, limited sample availabilities inhibited very long deposition times and, as a consequence, a high accumulation of matrix-isolated molecules on the spectroscopy window. In addition, the molecular compositions of the mixtures are prone to structural modifications upon laser vaporization. For the pellet measurements, the resolution of the spectrometer was set to 2 cm$^{-1}$, whereas for the MIS experiments, it was 1 cm$^{-1}$.

\subsection{Spectra measured in pellets} \label{sec_pellets}
\begin{figure*}\begin{center}
\epsscale{2.15} \plotone{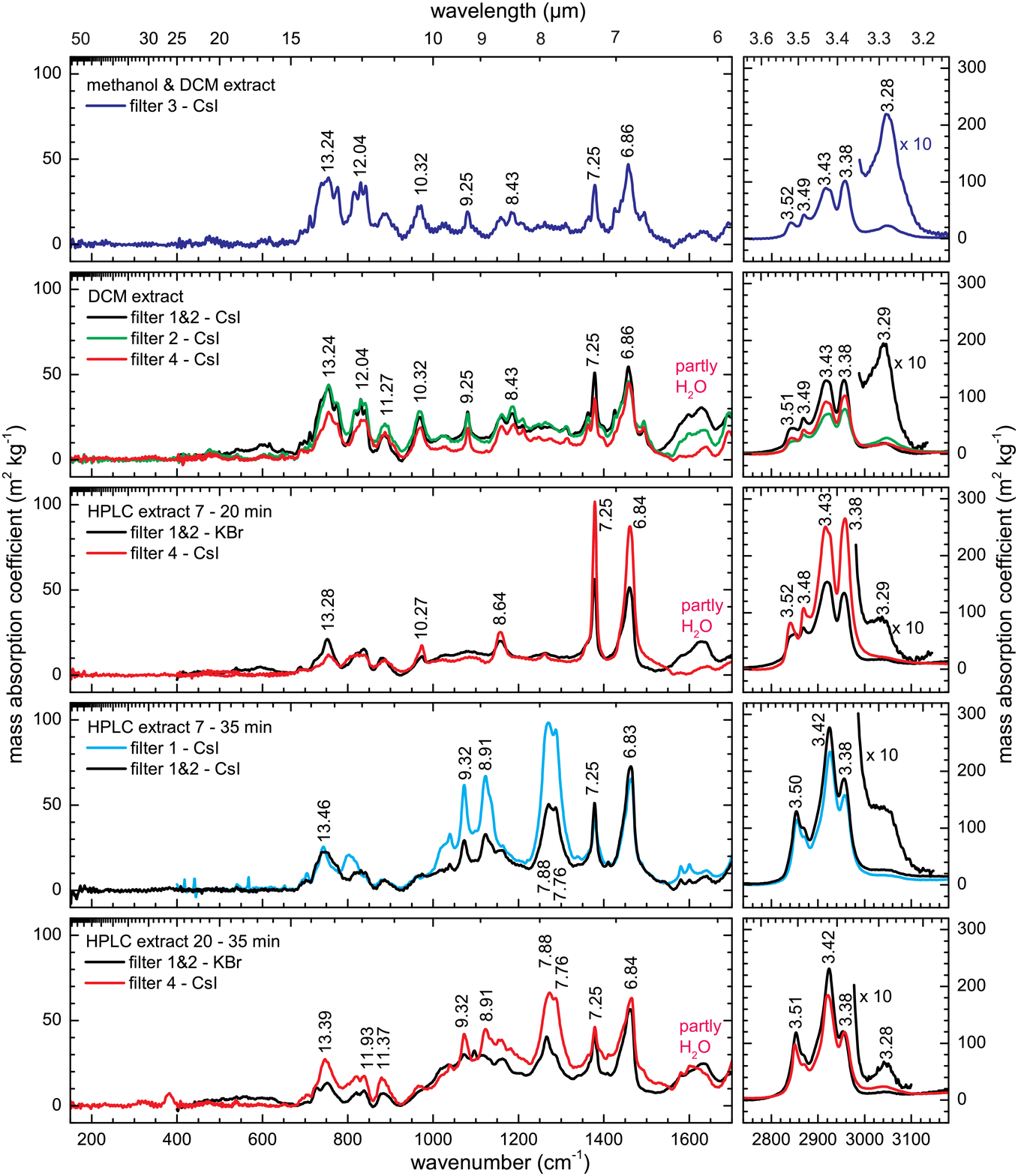} \caption{IR absorption spectra of PAH extracts from the laser pyrolysis embedded in CsI and KBr pellets at room temperature. The wavelength positions of the most prominent peaks are indicated. Note the different vertical scales of the left and right panels.} \label{fig_PAHs-pellets}
\end{center}\end{figure*}

The room temperature absorption spectra of the PAH extracts embedded in CsI and KBr pellets are presented in Fig. \ref{fig_PAHs-pellets}. All IR spectra were baseline-corrected to account for effects non-specific for the investigated samples (e.g., Christiansen effect). The most prominent band positions along with their interpretations are summarized in Table \ref{tab_5}. The assignments to vibrational modes of hydrogenated and methylated PAHs are mainly based on the DFT calculations discussed before. The broad band around 1630 cm$^{-1}$ (6.1 $\mu$m), which can be seen in some spectra, is in large part due to water. The bands between 700 and 1500 cm$^{-1}$ (14.3 $-$ 6.7 $\mu$m) are caused by bending and deformation modes in aromatic \includegraphics[scale=0.18]{arom-bond.eps}CH as well as aliphatic \includegraphics[scale=0.18]{doublebond.eps}CH$_2$ and \sbond CH$_3$ groups. Arising from compositional variations, the absolute band intensities and intensity ratios differ among the various samples. For instance, in comparison to the other vibrations, the CH out-of-plane bending modes between 700 and 920 cm$^{-1}$ (14.3 $-$ 10.9 $\mu$m) are stronger in those samples that contain more aromatic \includegraphics[scale=0.18]{arom-bond.eps}CH groups (i.e, the combined methanol/DCM and the DCM extract). According to the NMR analysis, the content of \includegraphics[scale=0.18]{doublebond.eps}CH$_2$ groups in the HPLC extract ``7 $-$ 35 min'' is higher than the content of \sbond CH$_3$ groups. This is reflected by the shape and substructure of the 3.4 $\mu$m absorption complex, which will be discussed shortly. Judging by the profile of the 3.4 $\mu$m feature, the sample ``20 $-$ 35 min'' is dominated by \includegraphics[scale=0.18]{doublebond.eps}CH$_2$ groups, too. The IR spectra of both substances display a broad band with two peaks at 1269 and 1288 cm$^{-1}$ (7.88 and 7.76 $\mu$m) as well as weaker peaks at 1073 and 1122 cm$^{-1}$ (9.32 and 8.91 $\mu$m), which are not present in the other samples.

In the following, we will concentrate on the CH stretching bands between 2800 and 3100 cm$^{-1}$ (3.57 $-$ 3.23 $\mu$m) as they are the strongest spectral features in our samples and because they are of higher relevance for comparison with astrophysical data. At first glance, the distinct intensity differences between the 3.3 $\mu$m band and the 3.4 $\mu$m complex are in qualitative agreement with the results from the NMR analysis summarized in Table \ref{tab_4}. Considering the intrinsic strengths of the stretching vibrations in CH$_x$ groups (Table \ref{tab_3}), the molecules in all samples exhibit strong coverage with excess hydrogens and methyl groups. From the quantitative laboratory spectra, we can also try to estimate the absorption cross sections per CH$_2$ and CH$_3$ groups. Taking into account the results from the NMR and MALDI-TOF analysis and assuming an average molecular mass of 350 amu with an average number of outer attachment sites for CH$_x$ groups of 14, the integrated mass absorption coefficients of the 3.4 $\mu$m bands transfer on average into an integrated cross section of about 10$^{-21}$ m$^2$ cm$^{-1}$ per CH$_2$/CH$_3$. The small difference between this value and the results obtained in Section \ref{sec_theoryIR} (Table \ref{tab_3}), which can be considered to be more accurate, may be justified by potential experimental error sources and simplifications.

Combining the knowledge from our previous DFT calculations, absorption measurements of individual molecules, and NMR investigations, we can now analyze the substructure of the 3.4 $\mu$m absorption complex in detail. Five to six separate peaks are apparent in the CH stretching region in the spectra of Fig. \ref{fig_PAHs-pellets} (right panels), corresponding to the one stretching mode in aromatic \includegraphics[scale=0.18]{arom-bond.eps}CH groups around 3045 cm$^{-1}$ (3.3 $\mu$m), the two stretching modes in \includegraphics[scale=0.18]{doublebond.eps}CH$_2$ groups, and the three stretching modes in \sbond CH$_3$ groups. According to the NMR analysis, the samples ``Methanol \& DCM extract'', ``DCM extract'', and ``HPLC extract 7$-$20 min'' contain more \sbond CH$_3$ than \includegraphics[scale=0.18]{doublebond.eps}CH$_2$ groups. This is reflected by enhanced peak heigths at 2870 cm$^{-1}$ (3.48 $\mu$m), 2917 cm$^{-1}$ (3.43 $\mu$m), and 2957 cm$^{-1}$ (3.38 $\mu$m). The peaks at 2847 cm$^{-1}$ (3.51 $\mu$m) and 2924 cm$^{-1}$ (3.42 $\mu$m), on the other hand, are emphasized in the spectrum of the \includegraphics[scale=0.18]{doublebond.eps}CH$_2$ dominated HPLC extract ``7$-$35 min'' as well as in the spectrum of the HPLC extract ``20$-$35 min'', for which no detailed NMR analysis is available, but which, apparently, also contains an excess of \includegraphics[scale=0.18]{doublebond.eps}CH$_2$ groups. Remember that the band positions given here are those of grains of PAHs embedded in halide salts at room temperature. The  energetic shifts against isolated molecules at low temperature are discussed in the following subsection. 

\begin{table*}
 \caption{Peak Assignments for the CsI Pellet Spectra Displayed in Fig. \ref{fig_PAHs-pellets}.}
\begin{tabular}{lll} \hline
wavenumber (cm$^{-1}$) & wavelength ($\mu$m) & vibrational mode \\ \hline\hline

700 $-$ 920 & 14.29 $-$ 10.87 & CH out-of-plane bending (aromatic component) \\

930 $-$ 1220 & 10.75 $-$ 8.20 & CH$_2$ out-of-plane bending, CC stretching, CO stretching \\

1260 $-$ 1380 & 7.94 $-$ 7.25 &  in-plane bending modes in \includegraphics[scale=0.18]{doublebond.eps}CH$_2$ and \includegraphics[scale=0.18]{arom-bond.eps}CH groups\\ 

1461 $\pm$ 3 & 6.84 $\pm$ 0.02 & \includegraphics[scale=0.18]{doublebond.eps}CH$_2$ and \sbond CH$_3$ deformation modes\\

1550 $-$ 1660 & 6.45 $-$ 6.02 & CC stretching and OH bending from H$_2$O impurities \\

1730 $\pm$ 1 & 5.78 & CO stretching (not in all samples) \\

2847 $\pm$ 7 & 3.51 $\pm$ 0.01 & aliphatic CH$_2$ stretching (``out-of-plane'' H) \\

2870 $\pm$ 2 & 3.48 $\pm$ 0.01 & aliphatic CH$_3$ stretching (``symmetric'') \\

2917 $\pm$ 1 & 3.43 & aliphatic CH$_3$ stretching (``asymmetric'') \\

2924 $\pm$ 1 & 3.42 & aliphatic CH$_2$ stretching (``in-plane'' H) \\

2957 $\pm$ 1 & 3.38 & aliphatic CH$_3$ stretching (``in-plane'' H) \\

3045 $\pm$ 5 & 3.28 $\pm$ 0.01 & aromatic CH stretching \\ \hline

\end{tabular}\\
\label{tab_5}
 \end{table*}

\subsection{Spectra measured in Ne matrices} \label{sec_matrix}
\begin{figure*}\begin{center}
\epsscale{2.15} \plotone{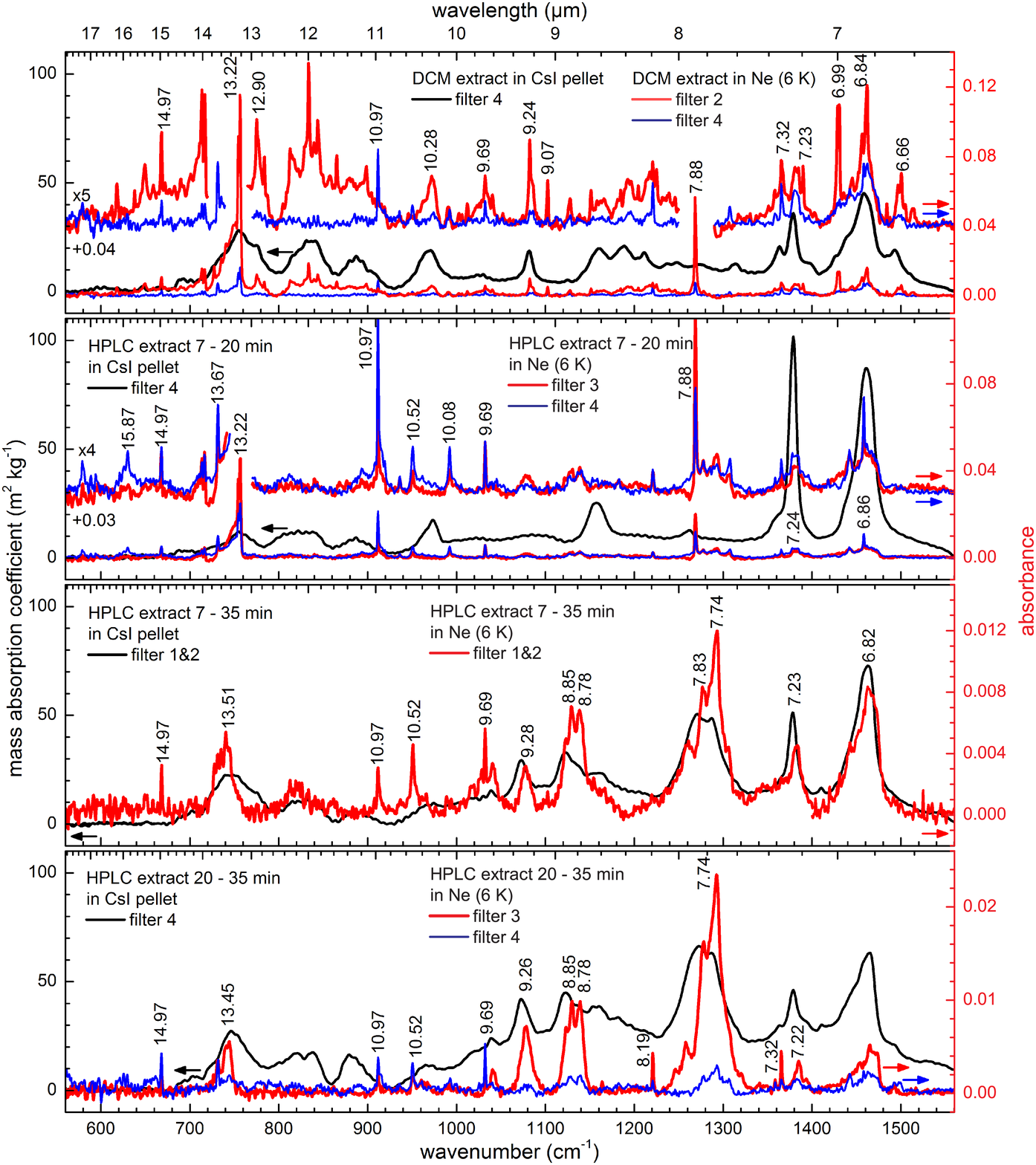} \caption{IR absorption spectra of PAH extracts from the laser pyrolysis embedded as grains in CsI pellets (black curves) in comparison with the spectra of isolated extract molecules at low temperature (6 K), which, prior to incorporation into the Ne matrix, were laser-evaporated (red and blue curves). The wavelength positions (in $\mu$m) of the most prominent peaks of the Ne matrix spectra are indicated.} \label{fig_MIS-pellets1}
\end{center}\end{figure*}

\begin{figure*}\begin{center}
\epsscale{2.15} \plotone{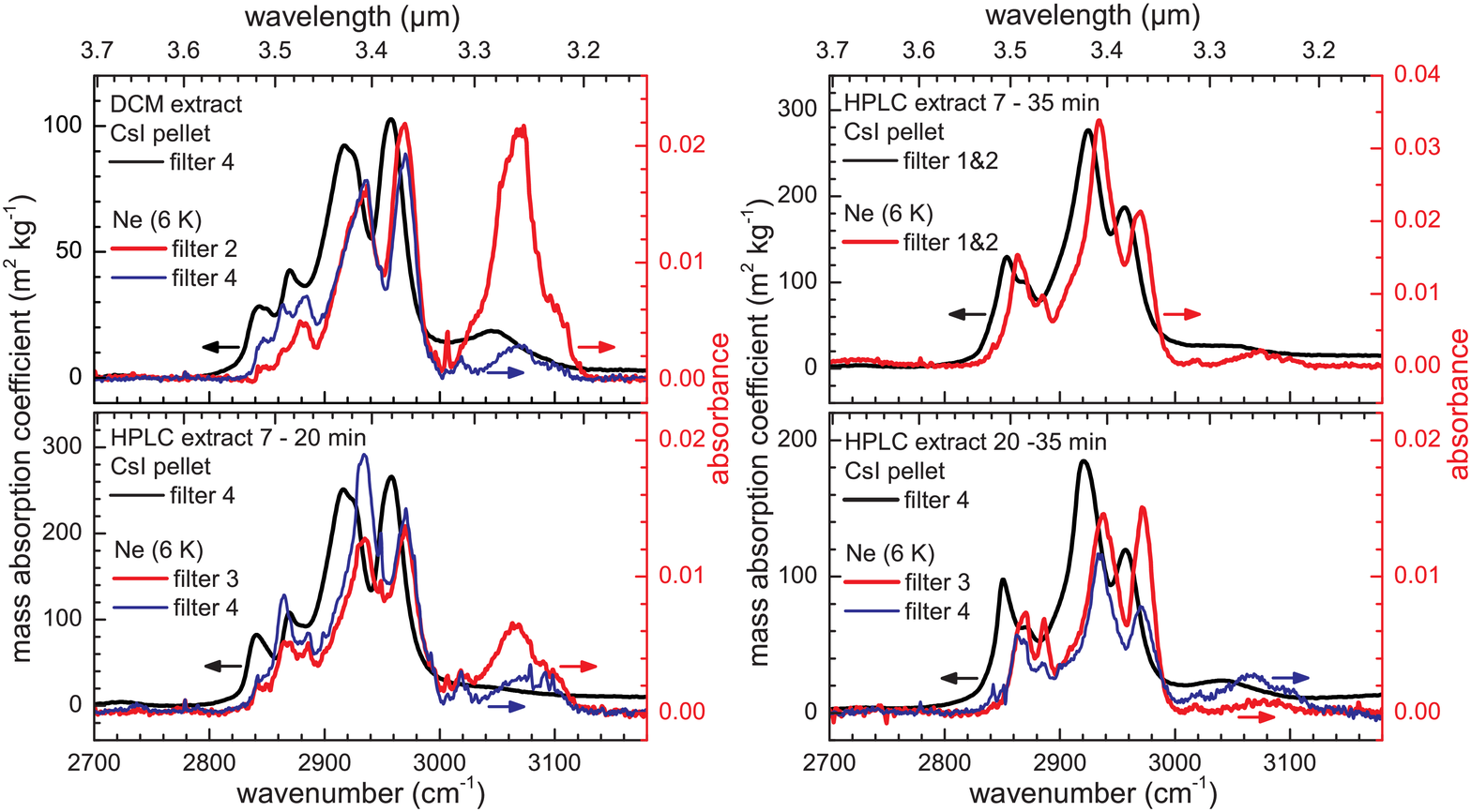} \caption{IR absorption spectra of PAH extracts from the laser pyrolysis embedded as grains in CsI pellets (black curves) in comparison with the spectra of isolated extract molecules at low temperature (6 K), which, prior to incorporation into the Ne matrix, were laser-evaporated (red and blue curves).} \label{fig_MIS-pellets2}
\end{center}\end{figure*}

\begin{figure}\begin{center}
\epsscale{1.0} \plotone{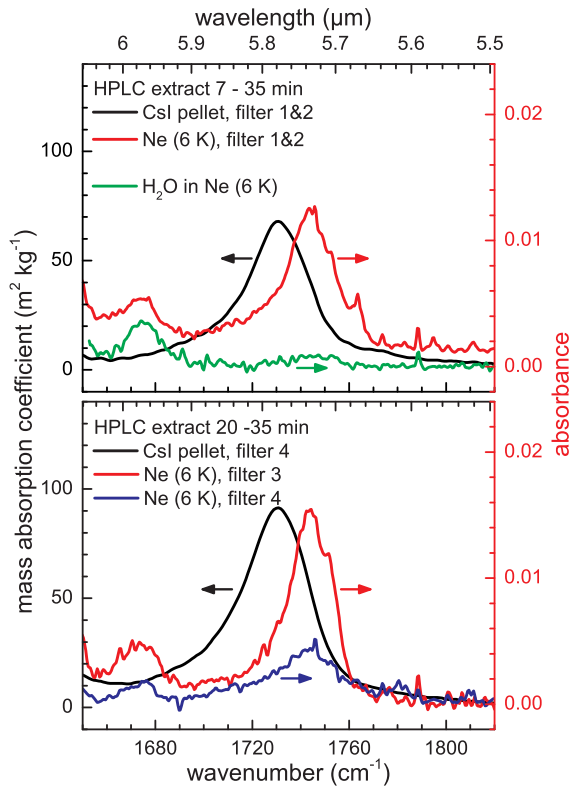} \caption{Stretching of C\dbond O bonds observed in two of the PAH extracts from the laser pyrolysis. Spectra of grains embedded in CsI pellets and of molecules isolated in solid Ne are displayed. For comparison, the upper panel shows the spectrum of an undoped Ne matrix with water impurities (green curve).} \label{fig_MIS-pellets3}
\end{center}\end{figure}

Figures \ref{fig_MIS-pellets1} and \ref{fig_MIS-pellets2} depict the low-temperature (6 K) absorption spectra of the matrix-isolated laser pyrolysis extracts in the mid-IR wavelength range (6 $-$ 18 $\mu$m and 3.1 $-$ 3.7 $\mu$m). The CO stretching bands, peaking at about 1744 cm$^{-1}$ (5.73 $\mu$m) in Ne matrix, observed for two of the extracts, are displayed in Fig. \ref{fig_MIS-pellets3}. For comparison, the upper panel of this figure contains the spectrum of an undoped Ne matrix. The small and broad feature at 1675 cm$^{-1}$ (5.97 $\mu$m) is caused by water isolated in the matrix. All low-temperature spectra are compared to their respective pellet measurements. The presented spectra were baseline-corrected for regular interference patterns, whose spacings (usually around 100 $-$ 150 cm$^{-1}$) and amplitudes (up to absorbance of 0.01 at 20 $\mu$m) depend on the wavelength and the thickness of the matrix. Comparably dimensioned (weak and broad) absorption features are very difficult to detect and might have been removed from the spectra along with the baseline correction procedure.

For each of the extracts measured by MIS, at least two experimental runs were performed, usually sampling materials from different filters. (The different filters sustained variable storage times in air, whereas the extracts from filter 1 and 4 are the oldest and newest samples, respectively.) Despite treating the samples as equal as possible, the absorption spectra of the DCM extract as well as the two HPLC extracts ``7$-$20 min'' and ``20$-$35 min'' display some variations among different experimental runs, as documented in the Figs. \ref{fig_MIS-pellets1} and \ref{fig_MIS-pellets2}. It seems that the laser vaporization procedure caused stronger modifications for those materials that were previously exposed to air for a longer time. The DCM extract (filter 2) and the HPLC extract ``7$-$20 min'' (filter 3) show clear signs of aromatization. The differences between the IR spectra of the matrix-isolated sample and the pellet spectra can predominantly be traced back to a conversion of aliphatic \includegraphics[scale=0.18]{doublebond.eps}CH$_2$ to aromatic \includegraphics[scale=0.18]{arom-bond.eps}CH groups, as well as a loss of \sbond CH$_3$ groups. This is especially obvious from the intensities of the peaks at 2935 cm$^{-1}$ (3.41 $\mu$m; \includegraphics[scale=0.18]{doublebond.eps}CH$_2$ and \sbond CH$_3$), 2969 cm$^{-1}$ (3.37 $\mu$m; \sbond CH$_3$), and 3068 cm$^{-1}$ (3.26 $\mu$m; \includegraphics[scale=0.18]{arom-bond.eps}CH). For the  HPLC extract ``20$-$35 min'' (filter 3), the main difference seems to concern the contents of the aliphatic groups alone. Besides the laser desorption process, this may also be explained by slight variations in the original compositions of the used samples.

Figure \ref{fig_MIS-pellets2} demonstrates the spectral differences between PAHs being locked up in grains at room temperature and being isolated in Ne matrix at 6 K. Compared to their room temperature counterparts, the aliphatic CH stretching bands of the matrix-isolated molecules are blue-shifted by approximately 10 $-$ 15 cm$^{-1}$. The shift of the weak and broad aromatic band at 3.3 $\mu$m is about 20 $-$ 30 cm$^{-1}$. The band widths of the stretching vibrations, on the other hand, are obviously not affected by the different experimental conditions, because large sets of molecules contribute, whereby each molecule has absorption bands with slightly varying energetic positions and intensities. 

The CO stretching bands, pictured in Fig. \ref{fig_MIS-pellets3}, were observed in the spectra of the two samples that contained more \includegraphics[scale=0.18]{doublebond.eps}CH$_2$ than \sbond CH$_3$ groups (i.e., the HPLC extracts ``7$-$35 min'' and ``20$-$35 min''). Their $^{13}$C-NMR spectra showed comparatively weak peaks from CO bonds, indicating the presence of only minor amounts of oxygen. However, because of a high intrinsic strength, the IR absorption of the CO stretching vibration is quite intense. Its intrinsic strength is about five to ten times higher than that of the aromatic \includegraphics[scale=0.18]{arom-bond.eps}CH stretching vibration \citep{pendleton02} and about as high as the combined aliphatic stretching vibrations in \includegraphics[scale=0.18]{doublebond.eps}CH$_2$ or \sbond CH$_3$ groups. The oxidization likely happened along with the sample treatment and storage under ambient conditions. Nevertheless, an oxidization during the laser pyrolysis, caused by trace amounts of water in the flow reactor, cannot be excluded either.

\section{Application to astrophysics} \label{sec_ISM}
\subsection{Stability of gas phase PAHs with excess hydrogenation and methyl groups} \label{sec_stability}
Physically bound PAHs can easily evaporate from a grain after absorption of a UV photon. \citet{jochims99} determined the photostability of several gas-phase PAHs by measuring the appearance energies of ionic photofragments. (Only dissociative photoionization processes were monitored.) It was found that a PAH carrying either a methyl group or a pair of dihydro groups had a lower photostability than the unsubstituted molecule. The dissociative photoionization process that requires the least energy separates one H atom from the parent molecule. The appearance energies observed for this process range from 11.85 eV for 1,2-dihydronaphthalene (C$_{10}$H$_{10}$) to 13.5 eV for 1-methylanthracene (C$_{15}$H$_{12}$) \citep{jochims99}. In methyl-substituted PAHs, the H atom comes from the methyl group unless the structure of the molecule is rearranged before the dissociation takes place. Regarding PAHs carrying dihydro groups, the H atom is released by one of these groups \citep[][and references therein]{jochims99}. While it was determined that normal PAHs were stable when exposed to the photon flux prevailing in H I regions, it was found that their methyl- and dihydro-substituted derivatives fragmented under the same conditions.

It is interesting to note that PAHs containing dihydro groups may play a role in the formation of interstellar H$_2$. The release of molecular hydrogen upon photodissociation of these species was suggested by \citet{jochims99}. The required energies for H$_2$ release during dissociative photoionization were on the order of 15 eV, a rather high energy considering the interstellar photon flux. However, dihydro-substituted PAHs can also fragment into neutral products. Recently, \citet{fu12} reported the formation of H$_2$ in the photodissociation of (neutral) H2Ant isolated in an Ar matrix. It occurred at energies of less than 5.5 eV. Molecular H$_2$ loss from protonated H2Phn was also reported by \citet{szczepanski11}.

\subsection{The electronic spectra of hydrogenated and methylated PAHs} \label{sec_UV}
In this subsection, applying theoretical methods only, we will evaluate how the electronic spectra of PAH mixtures are altered when H atoms or methyl groups are added to the molecular periphery. This issue may have direct astrochemical implications since the potential presence of such compounds in interstellar (or circumstellar) environments can be concluded from observational data not only in the IR, but also in the UV-visible. In the following, we will concentrate on the possible presence of PAHs in the (local) diffuse ISM, for which the extinction is known over a wide wavelength range. Unfortunately, there are no sight lines combining both IR and UV observations. Nevertheless, by means of quantitative estimates, we will test whether the spectra of methylated and hydrogenated PAHs are compatible with the astronomical mid-IR features (see the next subsection) as well as with the interstellar extinction curve in the UV. 

It was already shown that the absorption of a PAH blend, consisting of sufficiently different molecules, culminates in a broad bump in the UV \citep{cecchi-pestellini08, joblin92, malloci04, steglich10, steglich11}. The chromophore responsible for this feature is the aromatic plane that is present in every PAH. The size distribution of the aromatic planes and their average dimension determine the width and the position of the bump. In order to assign the whole interstellar 217.5 nm absorption feature to a mixture of PAHs, it was concluded that on average about $N_\text{C}$ = (90 $\pm$ 30) $\times$ 10$^{-6}$ $N_\text{H}$ carbon atoms have to be locked up in the aromatic planes, and that the average plane size is around 50 $-$ 60 C atoms, whereby grains of PAHs provided a better fit to the UV bump than isolated molecules \citep{steglich10}. The reader might want to know that, at least partly based on experimental results, other more or less well-defined molecular entities or chromophores were also suggested to be connected to the interstellar UV bump. These include naphthalene-based aggregates \citep{arnoult00, beegle97}, strongly dehydrogenated PAHs \citep{duley98, duley04, malloci08}, organic carbon and amorphous silicates as present in interplanetary dust particles \citep{bradley05}, pentagonal corner defects in polyhedral multilayer carbon nanoparticles \citep{pikhitsa05}, and fullerenes and carbon onions \citep{iglesias-groth2004, tomita02}.

\begin{figure}\begin{center}
\epsscale{1.0} \plotone{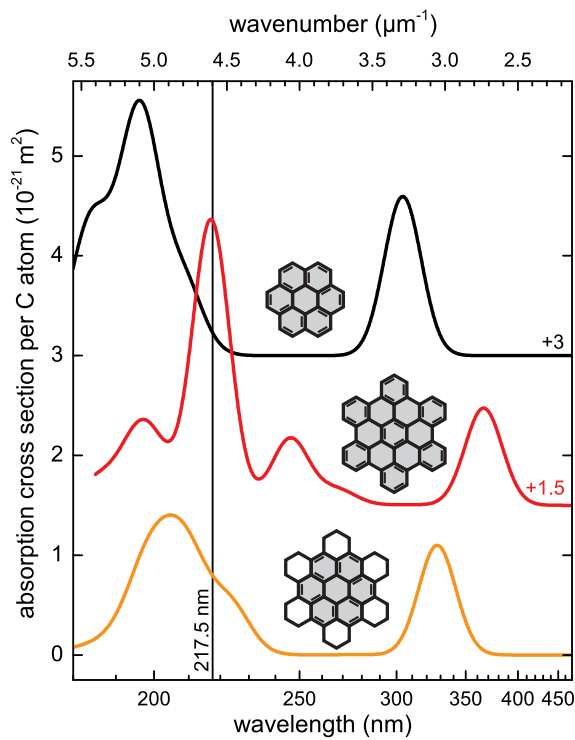} \caption{Theoretical electronic spectra of coronene (C$_{24}$H$_{12}$; top), hexabenzocoronene (C$_{42}$H$_{18}$; middle), and octadecahydrohexabenzocoronene (C$_{42}$H$_{36}$; bottom) calculated by TDDFT at the B3LYP/6-31+G(d) level of theory. Along with the chemical structures of the molecules, fully aromatic cycles are marked in gray.} \label{fig_hydrHBC}
\end{center}\end{figure}

We calculated the electronic spectra of hydrogenated and methylated PAHs applying time-dependent density functional theory (TDDFT). Before that, the molecular geometries were optimized and, to test for proper ground-state structures, the IR spectra were calculated using the B3LYP functional and the 6-31G basis set (like before in Section \ref{sec_indiv}). For the calculation of the electronic absorptions, we included polarization and diffuse functions, applying the 6-31+G(d) basis set \citep{frisch84}. The spectra were obtained by convoluting the transition energies and oscillator strengths of the different electronic transitions with Gaussian functions with full width at half maximum of 3000 cm$^{-1}$. Figure \ref{fig_hydrHBC} presents the computed spectra of the three molecules coronene (Cor; C$_{24}$H$_{12}$), hexabenzocoronene (HBC; C$_{42}$H$_{18}$), and octadecahydrohexabenzocoronene (H18HBC; C$_{42}$H$_{36}$). It demonstrates the dependence of the UV resonance on the size of the aromatic plane. HBC contains thirteen aromatic cycles. Its UV resonance is redshifted by 0.61 eV compared to Cor, which is composed of only seven rings. The largest molecule, H18HBC, is the peripherally fully hydrogenated derivative of HBC, but, like Cor, it only possesses seven aromatic rings. Consequently, its UV resonance is displaced back again to a shorter wavelength by about 0.34 eV compared to HBC. The hydrogenated PAH H18HBC can also be understood as Cor with additional aliphatic side groups, which are known to induce slight red shifts.

\begin{figure}\begin{center}
\epsscale{1.0} \plotone{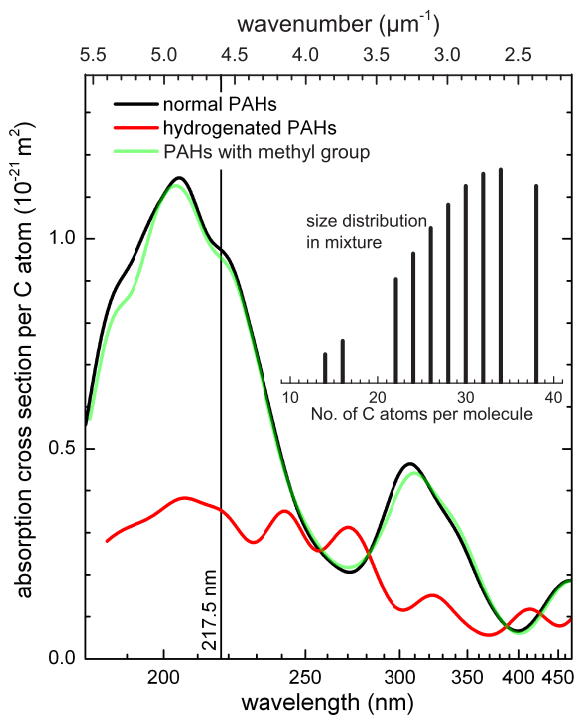} \caption{Theoretical electronic spectra of artificial molecular mixtures of normal PAHs, on their periphery fully hydrogenated PAHs, and PAHs with one methyl group. The spectra were calculated by TDDFT at the  B3LYP/6-31+G(d) level of theory. Inset: Size distribution of the molecular mixture. The chemical structures of the individual PAHs are displayed in Fig. \ref{fig_overviewPAHs}.} \label{fig_UV-hydroPAHmix}
\end{center}\end{figure}

In Section \ref{sec_theoryIR}, the calculated IR spectra of artificial blends of normal, hydrogenated, and methylated PAHs were presented. Now, we introduce the theoretical electronic spectra of those mixtures. For one of the size distributions we applied before (those with maximum at 34 C atoms), the theoretical electronic spectra are pictured in Fig. \ref{fig_UV-hydroPAHmix}. Despite a rather limited number of different molecular structures, a clear UV resonance around 205 nm is obvious for the normal PAHs. The spectrum of the PAHs with one additional methyl group is almost identical as the aromatic planes are basically unaltered. On the other side, the spectrum of the PAHs that are fully hydrogenated on their periphery does not exhibit a clear UV signature at all. This, however, is not very surprising because the maximum aromatic size in this mixture corresponds to Ant (three cycles), while most molecules have zero or only one aromatic ring.

To summarize, we can state that mixtures of hydrogenated and methylated PAHs also give rise to a strong UV bump, as long as the molecules exhibit an aromatic carbon network. The position of the bump is determined by the average size of the aromatic planes. Its intensity scales with the number of available $\pi$ electrons. The absorption cross section per $\pi$ electron is basically the same as in normal PAHs.

\subsection{Diffuse interstellar medium}
\begin{figure*}\begin{center}
\epsscale{2.15} \plotone{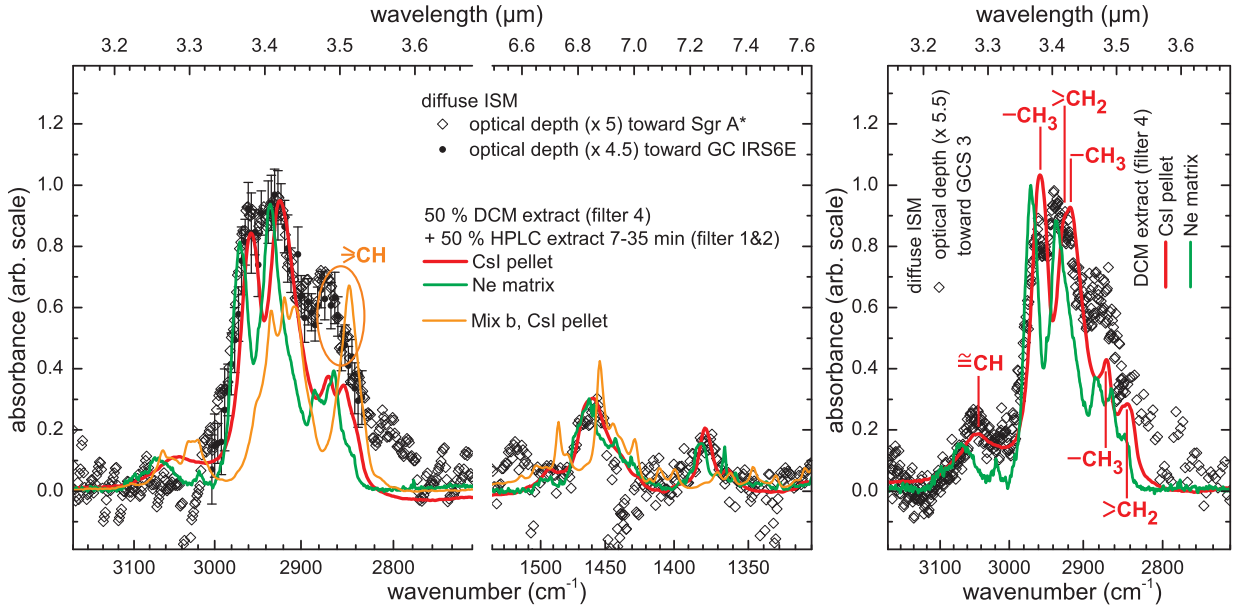} \caption{Laboratory spectra of hydrogenated and methylated PAHs in the solid phase compared with absorption features of the diffuse ISM observed toward Sgr A$^*$ \citep{chiar00}, GCS3 \citep{chiar00}, and GC IRS6E \citep{pendleton94}.} \label{fig_PAHs_GC}\end{center}\end{figure*}

We are going to test whether PAHs with methyl and methylene groups can explain the observed IR absorption features of the diffuse ISM. We start with a comparison of the band profiles and positions.
Figure \ref{fig_PAHs_GC} compares laboratory spectra of hydrogenated and methylated PAHs with the absorption bands of the diffuse ISM around 3.3, 3.4, 6.8, and 7.3 $\mu$m \citep{chiar00, pendleton94}. The shown samples are extracts from the laser pyrolysis, either in Ne matrix or in CsI pellet, and, in addition, the average spectrum of the individual PAHs with excess hydrogens on outer and inner C atoms displayed in Fig. \ref{fig_average_spectra} (Mix b).
The comparison with the ISM observations demonstrates a rather good overlap with the experimental data from the laser pyrolysis samples. Especially, the positions and relative intensities of the 6.8 and 7.3 $\mu$m features are well matched. The small variations in the subpeak positions of the 3.4 $\mu$m feature can likely be explained by different environmental conditions (temperature, isolated molecules vs. clusters) as can be seen from the differences between the pellet and Ne matrix spectra. Especially the central peak at 2925 cm$^{-1}$ (3.42 $\mu$m) seems to be better reproduced by the low-temperature spectra. At the same time, the positions of the subpeak at 2955 cm$^{-1}$ (3.38 $\mu$m) and of the 3.3 $\mu$m band might be slightly better matched by the pellet spectra, suggesting a stronger presence of clusters in the ISM. The differences are not very large, though, and the matrix-to-gas phase shifts might vary for the different functional groups. Probably, free-flying as well as clustered molecules at low temperature contribute to the astronomical IR features. However, the observations display a distinct excess of absorption around 2880 cm$^{-1}$ (3.47 $\mu$m), which hints at the presence of tertiary \includegraphics[scale=0.18]{triplebond.eps}CH groups (i.e., the hydrogenation of inner C atoms forming diamond-like structures).

\begin{figure}\begin{center}
\epsscale{1.0} \plotone{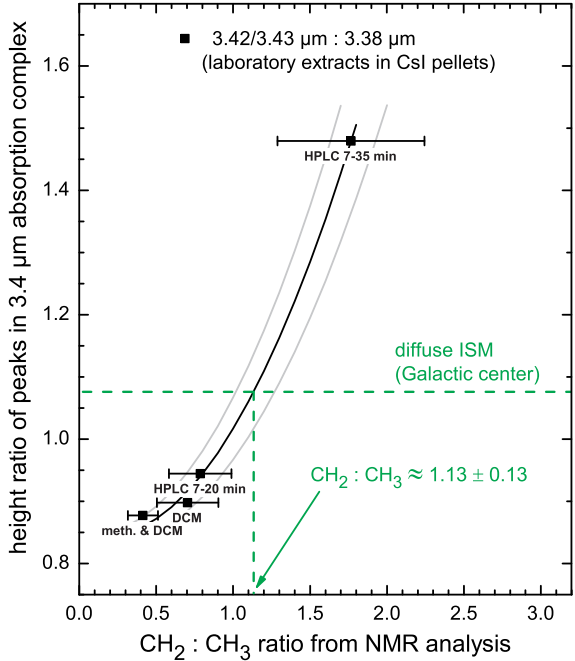} \caption{Approximate determination of the relative abundances of \protect\includegraphics[scale=0.18]{doublebond.eps}CH$_2$ and \sbond CH$_3$ groups in the diffuse ISM (see the text for an explanation).} \label{fig_height-ratios}
\end{center}\end{figure}

The above comparison illustrates that the diffuse ISM contains abundant amounts of methyl and methylene groups. The experimental results allow an estimation of the relative interstellar abundances of both functional groups. We tried to fit the sub-bands of the 3.4 $\mu$m feature of the experimental spectra to the five vibrational components and link the integrated areas to the NMR results. Unfortunately, due to the overlap of bands and unknown or unsystematic band profiles the results were not reliable enough. Instead, we chose to correlate only the peak heights of the bands at 3.42/3.43 $\mu$m and 3.38 $\mu$m with the experimental $N_{\text{CH2}}$$ / $$N_{\text{CH3}}$ ratio as determined from the NMR analysis. The result is pictured in Fig. \ref{fig_height-ratios}. For the purpose of interpolation, the data points were fitted to a second-order polynomial. By comparing with the approximate height ratio of the interstellar bands at 3.42 $\mu$m and 3.38 $\mu$m, we find that, in the diffuse ISM, the abundance ratio is about $N_{\text{CH2}}$$ / $$N_{\text{CH3}}$ = 1.13 $\pm$ 0.13. Additionally, tertiary \includegraphics[scale=0.18]{triplebond.eps}CH groups are likely responsible for the excess in the 3.47 $\mu$m feature. A very rough estimate, based only on the measured and calculated IR spectra of tetracosahydrocoronene (C$_{24}$H$_{36}$; see Section \ref{sec_HinnerC}), suggests that the $N_{\includegraphics[scale=0.12]{triplebond.eps}\text{CH}}$$ / $$N_{\text{CH2}}$ ratio is about one or even higher.

In the following, we determine upper limits for the absolute amounts of methyl and methylene groups that are needed in order to account for the complete interstellar 3.4 $\mu$m band. To be more precise, we will make use of the observed optical depth of the central peak at 3.42 $\mu$m, from which we can calculate the total quantity of both functional groups. For the local diffuse ISM, \citet{sandford95} obtained $A_v / \tau_{(3.4 \text{ } \mu m)}$ = 250 $\pm$ 40, where $A_v$ is the visual extinction and $\tau_{(3.4 \text{ } \mu m)}$ is the optical depth of the 3.42 $\mu$m peak. For the average galactic reddening of $R_v$ = 3.1, the column density of hydrogen $N_{\text{H}}$ is about $N_{\text{H}} / A_v$ $\approx$ 1.87 $\times$ 10$^{21}$ cm$^{-2}$ mag$^{-1}$ \citep{draine03}, from which $N_{\text{H}} / \tau_{(3.4 \text{ } \mu m)}$ = (4.68 $\pm$ 0.75) $\times$ 10$^{23}$ cm$^{-2}$ can be derived. The optical depth $\tau_{(3.4 \text{ } \mu m)}$ = $N_{\text{CH2+CH3}}$ $\times$ $\sigma_{(3.4 \text{ } \mu m)}$ is the product of the 3.42 $\mu$m absorption cross section $\sigma_{(3.4 \text{ } \mu m)}$ and the combined column densities of methyl and methylene groups $N_{\text{CH2+CH3}}$ (as both functional groups contribute to the peak). According to the results obtained in Section \ref{sec_cross-sect}, the peak cross section $\sigma_{(3.4 \text{ } \mu m)}$ amounts to 
\begin{equation}
\sigma_{(3.4 \text{ } \mu m)} = (7.9 \pm 2.4) \times 10^{-20} \text{ cm}^2 \text{ per CH}_2/\text{CH}_3,
\end{equation}
from which we obtain the combined amount of methyl and methylene groups that would be needed to account for the complete 3.4 $\mu$m band:
\begin{equation}
N_{\text{CH2+CH3}} / N_{\text{H}} = (2.7 \pm 1.3) \times 10^{-5}.
\end{equation}
In combination with our subpeak analysis from above, we can also estimate the individual abundances of the functional  groups. They amount to
\begin{equation}
N_{\text{CH2}} / N_{\text{H}} = (1.44 \pm 0.75) \times 10^{-5}
\end{equation}
and
\begin{equation}
N_{\text{CH3}} / N_{\text{H}} = (1.27 \pm 0.67) \times 10^{-5}.
\end{equation}
The additional amount of \includegraphics[scale=0.18]{triplebond.eps}CH groups is about equal to that of the \includegraphics[scale=0.18]{doublebond.eps}CH$_2$ groups:
\begin{equation}
N_{\includegraphics[scale=0.12]{triplebond.eps}\text{CH}} / N_{\text{H}} \approx 1.44 \times 10^{-5}.
\end{equation}
Aromatic \includegraphics[scale=0.18]{arom-bond.eps}CH groups are absent from many lines of sight. At most (in GCS 3; see right panel in Fig. \ref{fig_PAHs_GC}), they reach an integrated intensity that translates into $N_{\text{\includegraphics[scale=0.12]{arom-bond.eps}CH}} / N_{\text{CH2+CH3}}$ $\lesssim$ 0.39 $\pm$ 0.21, or
\begin{equation}
N_{\text{\includegraphics[scale=0.12]{arom-bond.eps}CH}} / N_{\text{H}} \lesssim (1.06 \pm 1.06) \times 10^{-5}.
\end{equation}
Be reminded that the calculated abundances are rather upper boundaries for the real values as they were essentially derived from experimental spectra of molecules in the solid phase. The CH stretching vibrations of gas phase molecules might well be two to four times stronger \citep{joblin94}, accordingly leading to lower abundances.

\begin{figure}\begin{center}
\epsscale{1.0} \plotone{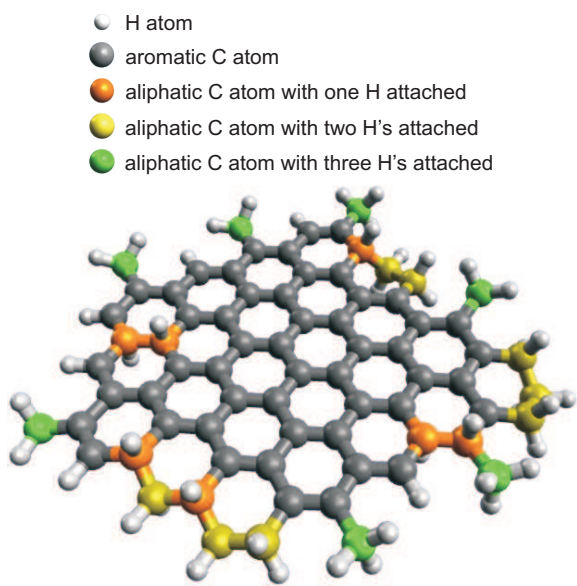} \caption{Inferred average molecular structure as present in the diffuse ISM. Such a structure, if mainly locked inside of grains, is compatible with the strengths, profiles, and positions of the UV bump at 217.5 nm as well as the IR absorption features around 3.3, 3.4, 6.8, and 7.3 $\mu$m.} \label{fig_prop-structure}
\end{center}\end{figure}

One might ask whether the estimated abundances of the individual functional groups are compatible with a molecular carrier (in the solid phase) that, at the same time, can also account for the complete 2175 \AA\ feature. Although this question might be difficult to answer, as there are no lines of sight combining observations in the IR and UV, we can at least check if the observed absorption strengths in both wavelength regimes would completely prevent such a carrier. In order to account for the whole UV bump strength (average value for $R_v$ = 3.1), $N_{\text{C}} / N_{\text{H}}$ = (9 $\pm$ 3) $\times$ 10$^{-5}$ carbon atoms have to be locked up in aromatic structures composed predominantly of 50 to 60 C atoms \citep{steglich10}. If we compare this value with the previous estimates, we find that $N_{\text{C}} / N_{\text{CH2+CH3}}$ = 3.3 $\pm$ 2.7, which is compatible with a large PAH (about 50 $-$ 60 aromatic C atoms) with excess hydrogens and methyl groups on its periphery. Figure \ref{fig_prop-structure} pictures the inferred average molecular structure, which, along with an almost unlimited number of variations, could be present in the diffuse ISM, either as free-flying unit or locked-up and interlinked in larger clusters or grains.

\subsection{Emission sources}
\begin{figure}\begin{center}
\epsscale{1.0} \plotone{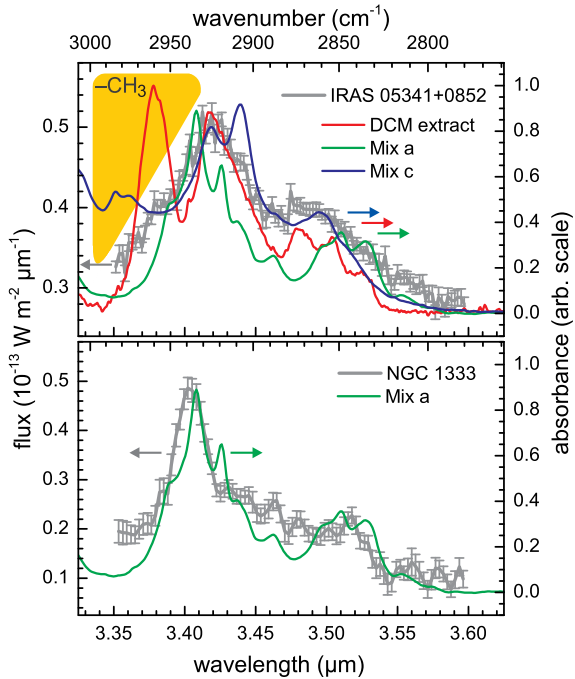} \caption{IR emission (gray lines; left vertical scale) observed from a proto-planetary nebula (top) and from a reflection nebula \citep[bottom; data from][]{joblin96} compared with laboratory spectra of hydrogenated and methylated PAHs (right vertical scale). The spectrum of the DCM extract was measured at low temperature in a Ne matrix and redshifted by 10 cm$^{-1}$. The CsI pellet spectra of Mix a (hydrogenated PAHs) and Mix c (methylated PAHs) were taken from Fig. \ref{fig_average_spectra}.} \label{fig_lab-emission}
\end{center}\end{figure}

As mentioned in the Introduction, some IR emission sources, especially proto-planetary nebulae, radiate in bands around 3.4 $\mu$m. In Fig. \ref{fig_lab-emission}, examples of such emission bands, observed from a proto-planetary nebula and a reflection nebula, are compared to laboratory spectra of hydrogenated and methylated PAHs in the solid phase. The laboratory spectrum of the DCM extract, which contains slightly more methyl than methylene groups, was measured with the MIS technique at 6 K. Due to anharmonicity, the IR emission bands of hot gas phase PAHs are shifted to the red compared to their absorption counterparts at low temperature. The redshift is usually on the order of 10 cm$^{-1}$ \citep{joblin95, wagner00}. Accordingly, the low-temperature spectrum of the DCM extract, as displayed in Fig. \ref{fig_lab-emission}, has been shifted by this amount. Because the room-temperature spectra of CsI pellets are already redshifted by approximately 10 cm$^{-1}$, we used them for comparison without further correction. The spectra of Mix a and Mix c were taken from Fig. \ref{fig_average_spectra}. While Mix a is solely composed of PAHs with excess hydrogenation on outer C atoms, Mix c contains only PAHs with methyl groups. The comparison with laboratory data demonstrates the absence of methylated PAHs in astronomical emission sources. Instead, the spectra of moderately hydrogenated PAHs match the observations quite well \citep[as already noted by][]{bernstein96, wagner00}. Likewise, the presence of tertiary \includegraphics[scale=0.18]{triplebond.eps}CH groups bound to PAHs can be ruled out because an excess of emission between 2850 and 2880 cm$^{-1}$ (3.51 $-$ 3.47 $\mu$m) is not observed.

\section{Conclusions} \label{sec_summary}
This contribution presented the IR absorption spectra of individual PAHs containing methyl and methylene groups as well as blends of methylated and hydrogenated PAHs produced by gas phase condensation. By combining quantitative laboratory spectra with DFT calculations, we were able to determine the absorption cross sections of hydrocarbon functional groups bound to PAHs in the solid phase. Additionally, supported by NMR and MALDI-TOF characterization techniques, the substructure of the aliphatic 3.4 $\mu$m absorption complex was analyzed in detail, and the six components (three from \sbond CH$_3$, two from \includegraphics[scale=0.18]{doublebond.eps}CH$_2$, and one from \includegraphics[scale=0.18]{triplebond.eps}CH) contributing to the subpeaks were revealed.

Armed with the knowledge obtained from the laboratory studies, the 3.4 $\mu$m absorption feature of the diffuse ISM was analyzed and upper limits for the total abundances of the individual hydrocarbon functional groups were determined. About equal amounts of the aliphatic groups \sbond CH$_3$, \includegraphics[scale=0.18]{doublebond.eps}CH$_2$, and \includegraphics[scale=0.18]{triplebond.eps}CH contribute to the absorption around 3.4 $\mu$m with maximum abundances of about $N_{\text{CH}x}$ $/$ $N_{\text{H}}$ $\approx$ 2 $\times$ 10$^{-5}$. If mainly gas phase PAHs are responsible for the 3.4 $\mu$m feature, the abundances, however, may be as low as $N_{\text{CH}x}$ $/$ $N_{\text{H}}$ $\approx$ 2 $\times$ 10$^{-6}$. Aromatic groups seem to be absent from some lines of sight, but can be almost as abundant as each of the aliphatic components in other directions ($N_{\text{\includegraphics[scale=0.12]{arom-bond.eps}CH}}$ $/$ $N_{\text{H}}$ $\lesssim$ 2 $\times$ 10$^{-5}$). Therefore, we conclude that, despite rather low binding energies, PAHs in the diffuse ISM are heavily decorated with excess hydrogens. Observations of astronomical IR emission sources, on the other hand, illustrate that the radiation from a nearby UV-visible source can easily strip away the excess hydrogenation, leaving almost purely aromatic molecules. At best, PAHs with methylene groups (i.e., only \includegraphics[scale=0.18]{doublebond.eps}CH$_2$, not \includegraphics[scale=0.18]{triplebond.eps}CH or \sbond CH$_3$) are present in some of those objects, especially in proto-planetary nebulae.

The spectral properties of the interstellar 3.4 $\mu$m absorption complex are not specific enough to allow a definite statement about the physical or chemical conditions of its carriers. Hydrogenated and methylated PAHs, either as free-flying units or locked-up and interlinked within nanoparticles or grains, likely contribute. The simultaneous presence of other molecular entities, containing functional hydrocarbon groups, which are likewise compatible with the subfeatures and spectroscopic constraints of the 3.4 $\mu$m complex, cannot be ruled out. However, the average molecular structure that we proposed to be abundantly present in the diffuse ISM, composed of an aromatic core with a mainly aliphatic rim, is also compatible with observations (and abundance constraints) in the UV. Such molecular units can account not only for the absorption bands in the IR, but also for the UV bump at 217.5 nm and the smooth extinction curve, that is observed for wavelengths shorter than 400 nm \citep{steglich10, steglich12}.

\acknowledgments
Part of this work was supported by the Deutsche Forschungsgemeinschaft, DFG project number Hu 474/21-2. We are grateful to J. Chiar, who kindly provided the data from her publication, H. Mutschke for providing access to the FTIR spectrometer, and G. Rouill\'e for constructive feedback. Technical assistance came from G. Born and S. T\"urk.

\end{document}